\documentclass[10pt, conference, compsocconf]{IEEEtran}
\ifCLASSINFOpdf
\else
\fi

\usepackage[usenames,dvipsnames]{xcolor}
\usepackage{amsmath,amsthm}
\usepackage{bm}
\usepackage{amssymb}
\usepackage{mathabx}
\usepackage{dsfont}
\usepackage{graphicx}
\usepackage{epstopdf}
\usepackage[noadjust]{cite}
\usepackage{etoolbox}
\usepackage{filecontents}
\usepackage{tikz}
\usepackage[ruled,vlined]{algorithm2e}
\usepackage{subcaption}

\begin{document}
%
\title{Re-thinking EEG-based non-invasive brain interfaces: modeling and analysis}

\author{\IEEEauthorblockN{Gaurav~Gupta\IEEEauthorrefmark{1},
S\'ergio~Pequito\IEEEauthorrefmark{2} and
Paul~Bogdan\IEEEauthorrefmark{1}
\IEEEauthorblockA{\IEEEauthorrefmark{1}Ming Hsieh Department of Electrical Engineering, University of Southern California, Los Angeles, CA, USA\\
Email: \{ggaurav, pbogdan\}@usc.edu}
\IEEEauthorblockA{\IEEEauthorrefmark{2}Department of Industrial and Systems Engineering, Rensselaer Polytechnic Institute, Troy, NY, USA\\
Email: goncas@rpi.edu}
}}


\maketitle

\begin{abstract}
Brain interfaces are cyber-physical systems that aim to harvest information from the (physical) brain through sensing mechanisms, extract information about the underlying processes, and decide/actuate accordingly. Nonetheless, the brain interfaces are still in their infancy, but reaching to their maturity quickly as several initiatives are released to push forward their development (e.g., NeuraLink by Elon Musk and `typing-by-brain' by Facebook). This has motivated us to revisit the design of EEG-based non-invasive brain interfaces. Specifically, current methodologies entail a highly skilled neuro-functional approach and evidence-based \emph{a priori} knowledge about specific signal features and their interpretation from a neuro-physiological point of view. Hereafter, we propose to demystify such approaches, as we propose to leverage new time-varying complex network models that equip us with a fractal dynamical characterization of the underlying processes. Subsequently, the parameters of the proposed complex network models can be explained from a system's perspective, and, consecutively, used for classification using machine learning algorithms and/or actuation laws determined using control system's theory. Besides, the proposed system identification methods and techniques have computational complexities comparable with those currently used in EEG-based brain interfaces, which enable comparable online performances. Furthermore, we foresee that the proposed models and approaches are also valid using other invasive and non-invasive technologies. Finally, we illustrate and experimentally evaluate this approach on real EEG-datasets to assess and validate the proposed methodology. The classification accuracies are high even on having less number of training samples.

\end{abstract}


%
\IEEEpeerreviewmaketitle

%
%
\section{Introduction}
\label{sec:intro}
We have recently testimony a renewed interest in invasive and non-invasive brain interfaces. Elon Musk has released the NeuraLink initiative~\cite{neuralink} that aims to develop devices and mechanisms to interact with the brain in a symbiotic fashion, thus merging the artificial intelligence with the human brain. The potential is enormous since it would ideally present a leap in our understanding of the brain, and an unseen enhancement of its functionality. Alternatively, Facebook just announced the `Typing-by-Brain' project~\cite{facebookTyping} that gathered a team of $60$ researchers whose target is to be capable of writing 100 words per minute that contrasts with the state-of-the-art of $0.3$ to $0.82$ words per minute assuming an average of $5$ letters per word. Towards this goal, Facebook has invested in developing new non-invasive optical imaging technology that is five times faster and portable with respect to the one available on the market and would possess increased spatial and temporal resolution. Nonetheless, these are just some of the initiatives among others by some big Silicon Valley players that want to commercialize brain technologies~\cite{siliconBrainTech}.

Part of the motivation for the `hype' in the use of neurotechnologies -- both invasive and non-invasive brain interfaces -- is due to their promising application domains~\cite{wolpaw2012brain}: (\emph{i})~\emph{replace}, i.e., the interaction of the brain with a wheelchair or a prosthetic device to replace a permanent functionality loss, (\emph{ii}) \emph{restore}, i.e.,  to bring to its normal use some reversible loss of functionality such as walking after a severe car accident or limb movement after a stroke, (\emph{iii}) \emph{enhance}, i.e.,  to enable to outperform in a specific function or task, as for instance an alert system to drive for long periods of time while attention is up; and (\emph{iv}) \emph{supplement} as in equipping one with extra functionality, as a third arm to be used during surgery. Notwithstanding, these are just some of the (known) potential uses of neurotechnology.

Despite the developments and promise of future applications of brain interfaces (some of which we cannot currently conceive), we believe that current approaches to both invasive and non-invasive brain interfaces can greatly benefit from cyber-physical systems~(CPS) oriented approaches and tools to increase their efficacy and resilience. Hereafter, we propose to focus on non-invasive technology relying on electroencephalogram (EEG) and revisit it through a CPS lens. Moreover, we believe that the proposed methodology can be easily applicable to other technologies, e.g., electromagnetic fields (magnetoencephalography~(MEG)~\cite{mellinger2007meg}, and the  hemodynamic responses associated to neural activity, e.g. functional magnetic resonance imaging~(fMRI)~\cite{sitaram2007fmri,weiskopf2004principles}, and functional
near-infrared spectroscopy~(fNIRS)~\cite{coyle2007brain}). Nonetheless, these technologies present several drawbacks compared to EEG, e.g., cost, scalability, and user comfort, which lead us to focus on EEG-based technologies. Similar argument can be applied in the context of non-invasive versus invasive technologies that require patient surgery.

\begin{figure*}
\centering
\begin{tikzpicture}[scale = 1.4]

\node[anchor=south west,inner sep=0] at (0,0) {\includegraphics*[viewport=0 0 850 310, width = 7in, height = 2.7in]{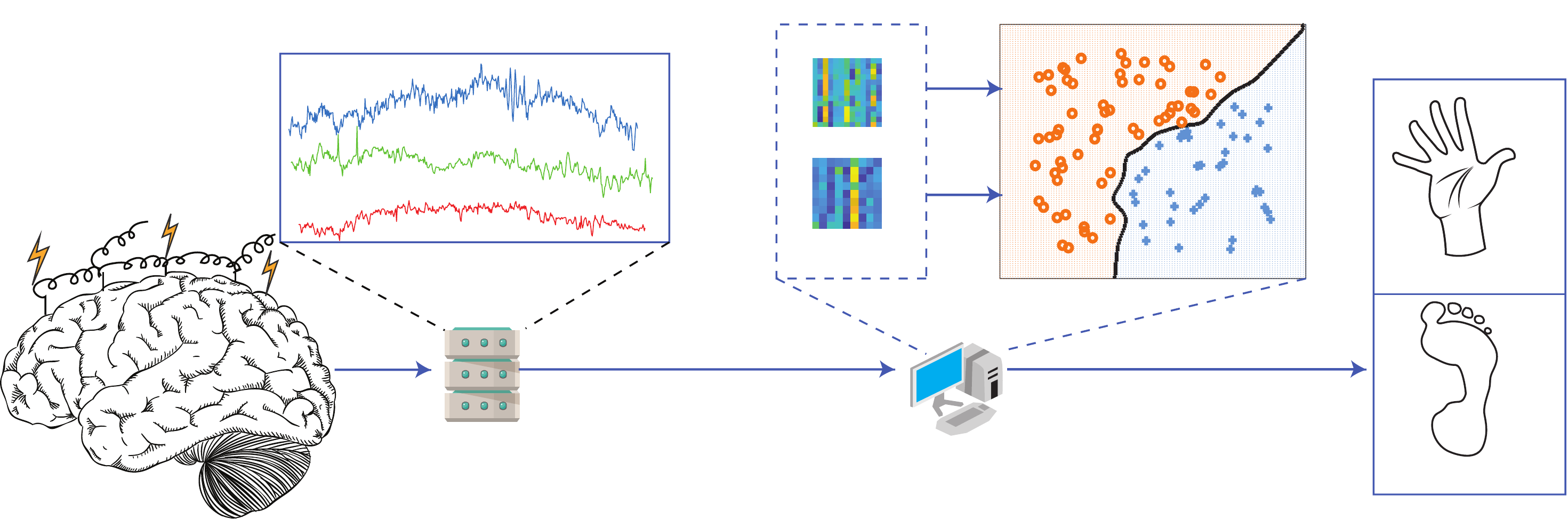}};

\node[anchor=north west] at (0.3,3) {EEG channels};
\node[anchor=north west] at (2.8,4.3) {EEG time series};
\node[anchor=north west] at (6.4,4.7) {$\substack{\text{feature} \\ \text{extraction}}$};
\node[anchor=north west] at (8.5,4.7) {$\substack{\text{machine learning} \\ \text{classification}}$};
\node[anchor=north west] at (11.3,4.2) {$\substack{\text{predicted} \\ \text{motor task}}$};
\node[anchor=north west] at (6.5,0.7) {$\substack{\text{real time processing of time series} \\ \text{for machine learning classification}}$};

\end{tikzpicture}

\caption{A systematic process flow of the Brain interface. The imagined motor movements of the subject are captured in the form of EEG time series which are then fed to the computational unit. A fractional-order dynamics based complex network model is estimated for the time series and the model parameters are used as features for machine learning classification. The output of classifier predicts various motor movements with certain confidence.}
\label{fig:bciProcess}
\end{figure*}
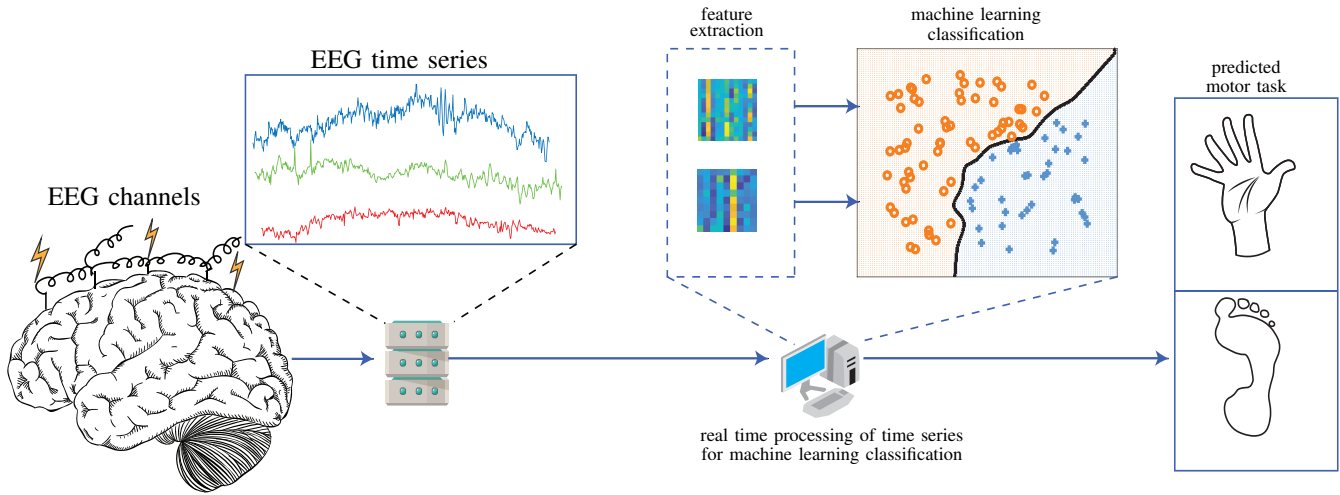

Traditional approach to EEG neuroadaptive technology consists of  proceeding through the following steps~\cite{wolpaw2002brain,lotte2014tutorial}: (\emph{a}) signal acquisition (in our case, measurement of the EEG time-series); (\emph{b}) signal processing (e.g., filtering with respect to known error sources); (\emph{c}) feature extraction (i.e., an artificial construction of the signal that aims to capture quantities of interest); (\emph{d}) feature translation (i.e., classification of the signal according to some plausible neurophysiological hypothesis); and (\emph{e}) decision making (e.g., provide instructions to the computer to move a cursor, or a wheelchair to move forward) -- see also Figure\,\ref{fig:bciProcess} for an overview diagram. 

In this paper, we propose to merge steps (b) and (c) motivated by the fact that these often accounts for spatial or temporal properties, and are only artificially combined in a later phase of the pipeline, i.e., at step (d) of feature translation. First, we construct a time-varying complex network where the activity of nodes (or EEG signal) have long-range memory and the edges accounts for inter-node spatial dependence. Thus, we argue that the previous approach discards several spatial-temporal properties that can be weighted for signal processing and feature extraction phases. In other words, current EEG brain interfaces require one to have an understanding of the different regions of the brain responsible. For instance, regions for motor or visual actions, as well as artificial frequency bands that are believed to be more significant for a specific action (also known as \emph{evidence-based}). Besides, one needs to understand and anticipate the most likely causes noise/artifacts in the EEG data collected and filter out entire frequency bands, which possibly compromises phenomena of interest not being available for post-processing. Instead, we propose a modeling capability to enable the modeling of long-range memory time-series that at the same time accounts for unknown stimuli, e.g., artifacts or inter-region communication.


\subsection{Related Work and Novel Contributions}

We put forward that the ability to properly model the EEG time-series with complex network models, that account for realistic setups, can enable the brain interfaces methods to get transformed from detectors to decoders. In other words, we do not want to solely look for the existence of a peak of activity in a given band that is believed to be associated with a specific action. But we want to decompose the bunch of signals into different features, i.e., parameters of our complex network model, that are interpretable. Thus, allowing us to understand how different regions communicate during a specific action/task by representing them as nodes of the complex network and estimating the dependence via coupling matrix. The node activities are assumed to be evolving at different time-scales and in the general setup of presence of external stimuli which could either be unwanted noise or external process driving the system. In engineering, this will enable us to depart from a skill dependent situation to general context analysis, which will enhance the resilience of the approaches for practical nonsurgical brain interfaces. Besides, it will equip bioengineers, neuroscientists, and physicians with an exploratory tool to pursue new technologies for neuro-related diagnostics and treatments, as well as neuro-enhancement.

The proposed approach departs from those available in the literature, see~\cite{wolpaw2002brain,lotte2014tutorial,wolpaw2012brain} and references therein. In fact, to the best of authors' knowledge, in the context of noninvasive EEG-based technologies,~\cite{brodu2012exploring} is the only existing work that explores fractional-order models in the context of single-channel analysis, which contrasts with the spatiotemporal modeled leveraged in this paper that copes with multi-channel analysis. Furthermore, the methodology presented in this paper also accommodate unknown stimuli~\cite{gaurav2017acc}. For which efficient algorithms are proposed and leveraged hereafter to simultaneously retrieve the best model that conforms with unknown stimuli, and separating the unknown stimuli from the time-series associated with brain activity.  Our methods are as computationally efficient and stable as least-squares and spectral analysis methods used in a spatial and temporal analysis, respectively; thus, suitable for online implementation in nonsurgical brain interfaces.


The main contributions of the present paper are those of leveraging some of the recently proposed methods to develop new modeling capabilities for the EEG based neuro-wearables that are capable of enhancing the signal quality and decision-making. Furthermore, the parametric description of these models provides us with new features that are biologically motivated and easier to translate in the context of brain function associated with a signal characterization, and free of signal artifacts. Thus, making the brain-related activity interpretable, which leads to resilient and functional nonsurgical brain interfaces.

\subsection{Paper Organization}

The remaining of the paper is organized as follows. Section~\ref{sec:probForm} introduces the complex network model considered in this paper and the main problem studied in this manuscript. Also we will see the description of the employed method for feature selection and then classification techniques. In Section~\ref{sec:experi}, we present an elaborated study on the different datasets taken from the BCI competition \cite{blankertz}.

\section{Re-thinking EEG-based non-invasive \\ brain interfaces}
\label{sec:probForm}

Brain interfaces aim to address the following problem.

\emph{Is it possible to classify a specific cognitive state, e.g., motor task or its imagination, by using measurements collected with a specific sensing technology that harvest information about brain activity?}

In the current manuscript, we revisit this problem in the context of brain-computer interfaces (BCI), when dealing with EEG-based noninvasive brain interfaces. Towards this goal, we review the currently adopted procedures for solving this problem (see Figure\,\ref{fig:bciProcess} for an overview), and propose a systems' perspective that enables to enhance the BCI reliability and resilience. Therefore, in Section\,\ref{ssec:eegOverview} we provide a brief overview of the EEG-based technology and the connection with the brain-areas' function associated with studies conducted in the past. Next, in Section~\ref{ssec:sysMod}, we introduce the spatiotemporal fractional model under unknown stimuli. This will be the core of the proposed approach in this manuscript to retrieve new features for classification, and, subsequently, enhancing brain interfaces capabilities. In Section~\ref{ssec:sysIdenti}, we describe how to determine the system model's parameters, and in Section~\ref{ssec:featureClassi} we describe how to interpret them for the classification task.

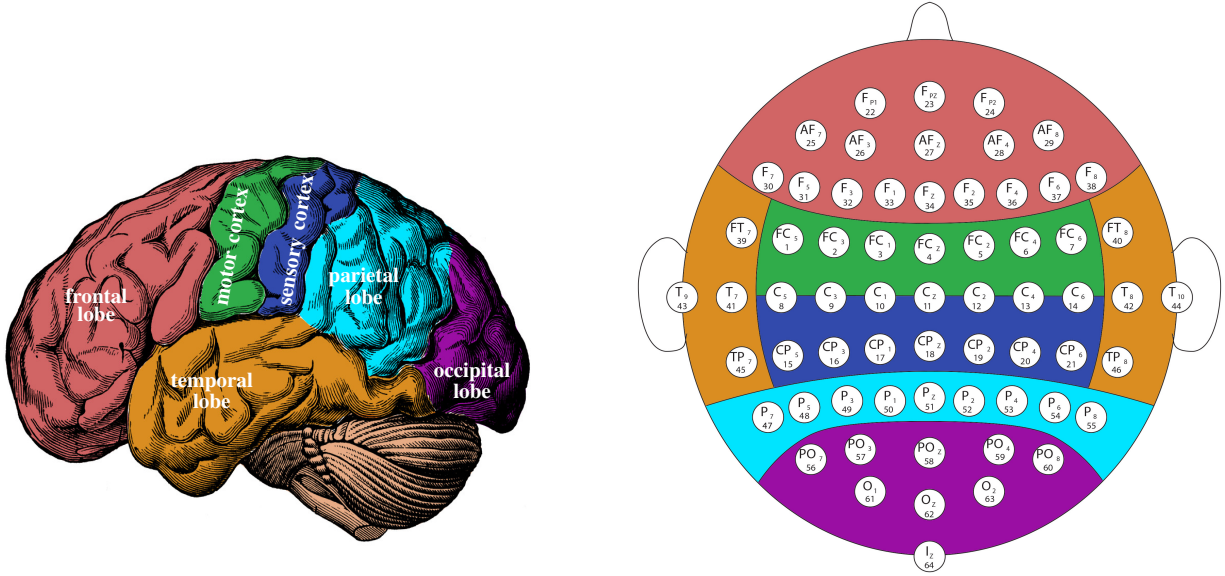
\begin{figure*}[!h]
\centering
\begin{tikzpicture}[scale = 1.4]

\node[anchor=south west,inner sep=0] at (0,0) {\includegraphics*[viewport=0 0 1300 730, width = 3.5in, height = 2.2in]{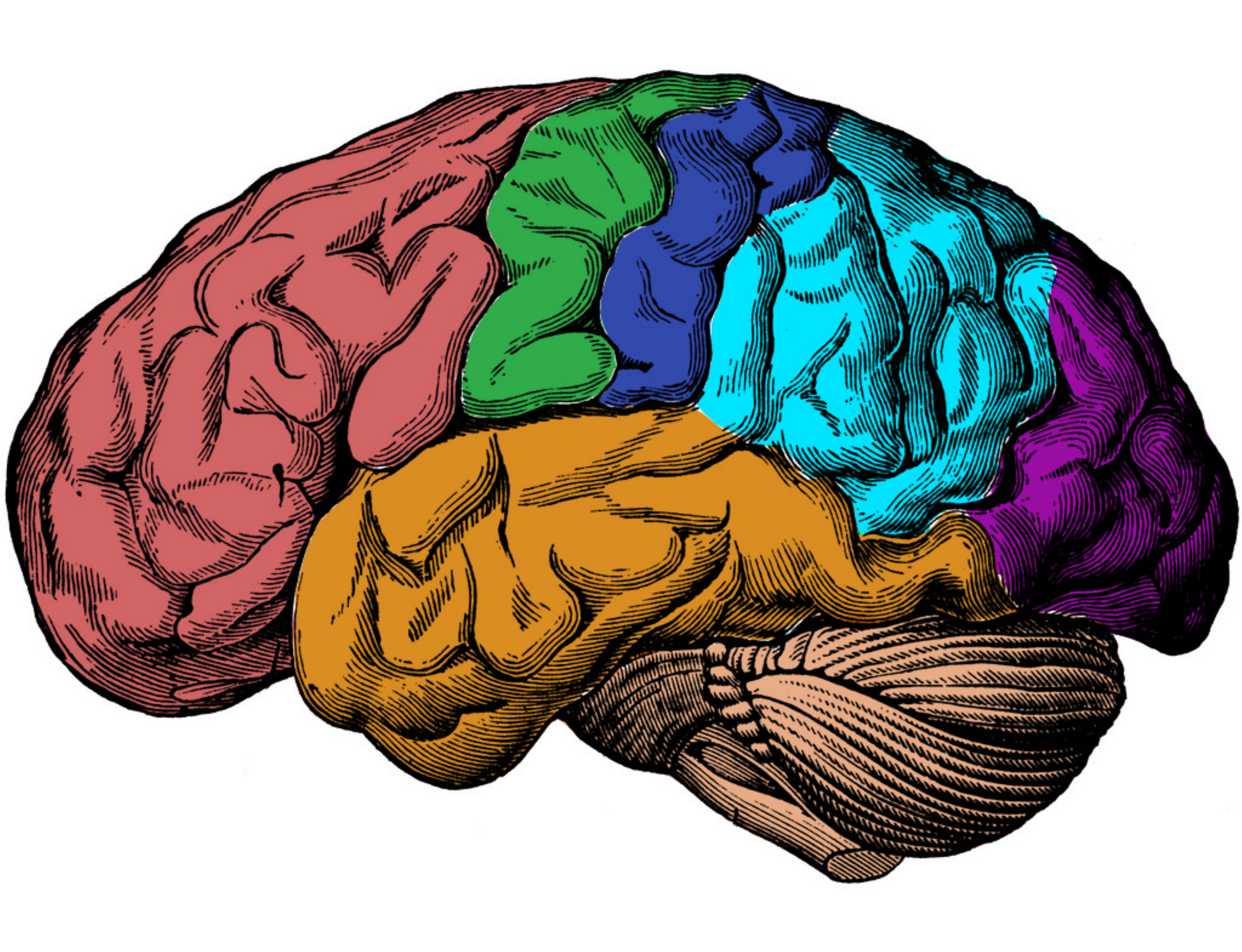}};

\node[anchor=south west,inner sep=0] at (6,0) {\includegraphics*[viewport=0 0 510 502, width = 3.1in, height = 3in]{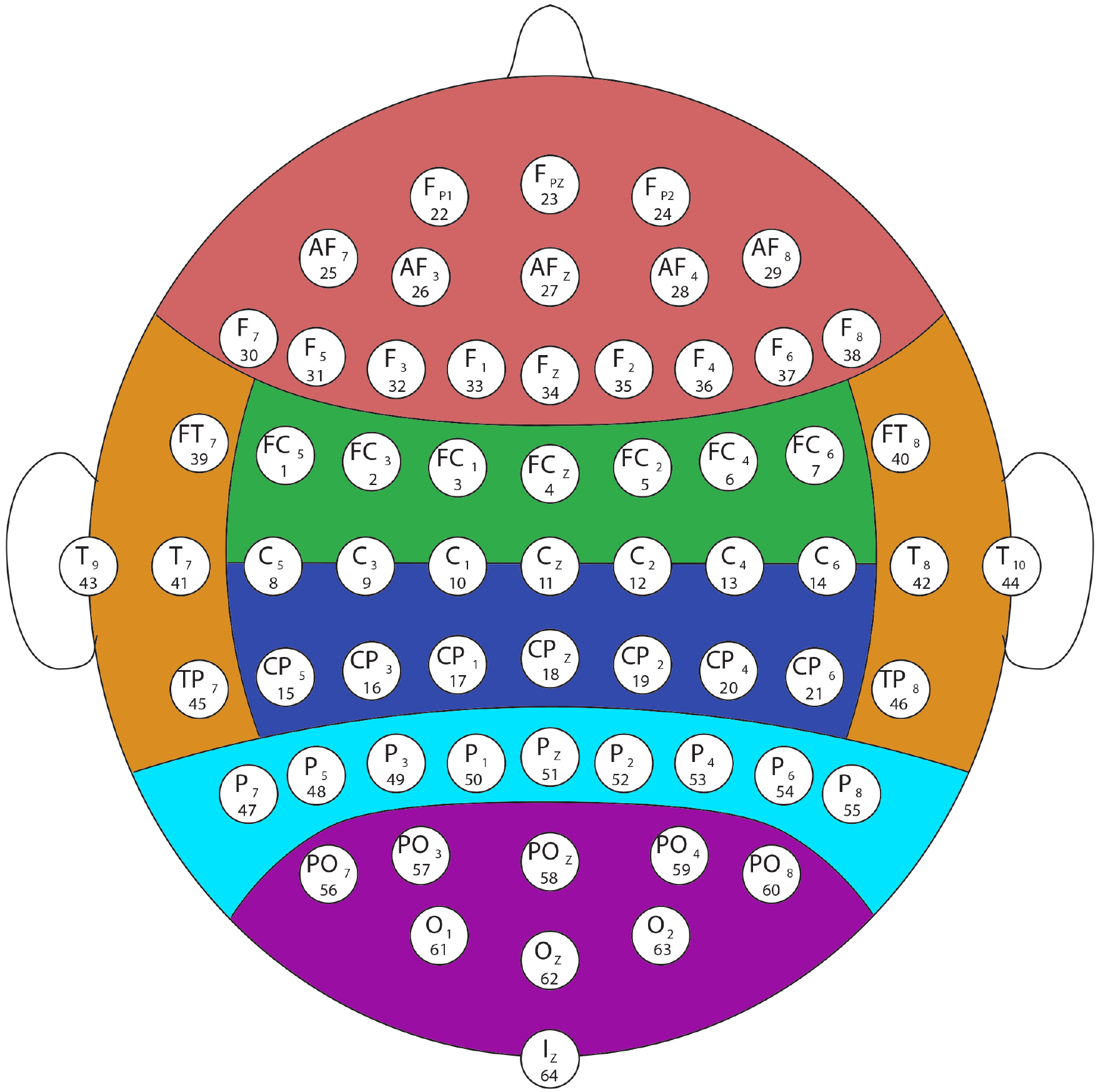}};

\node[anchor=north west,color=white] at (0.5,2.8) {$\large{\substack{\textbf{frontal} \\ \textbf{lobe}}}$};

\node[anchor=north west,color=white] at (1.5,2) {$\large{\substack{\textbf{temporal} \\ \textbf{lobe}}}$};

\node[anchor=north west,color=white,rotate=80] at (1.9,2.5) {\footnotesize
{\textbf{motor cortex}}};

\node[anchor=north west,color=white,rotate=80] at (2.5,2.5) {\footnotesize
{\textbf{sensory cortex}}};

\node[anchor=north west,color=white] at (3.0,3) {$\large{\substack{\textbf{parietal} \\ \textbf{lobe}}}$};

\node[anchor=north west,color=white] at (4.0,2.1) {$\large{\substack{\textbf{occipital} \\ \textbf{lobe}}}$};

\end{tikzpicture}
\caption{Description of the brain functional regions and their corresponding location with respect to the EEG sensor cap.}
\label{fig:EEGmontage}
\end{figure*}

\subsection{EEG-based Technology for Brain Interfaces: a brief overview}
\label{ssec:eegOverview}

EEG enables the electrophysiological monitoring of space-averaged synaptic source activity from millions of neurons occurring at the neocortex level. The EEG signals have a poor spatial resolution but high temporal resolution, since the electrical activity generated at the ensemble of neurons level arrives at the recording sites within milliseconds. The electrodes (i.e., sensor) are placed over an area of the brain of interest, being the most common the visual, motor, sensory, and pre-frontal cortices. Usually, they follow standard \emph{montages} -- the International 10-20 system is depicted in Figure\,\ref{fig:EEGmontage}.

Most of the activity captured by the EEG electrodes is due to the interactions between inhibitory interneurons and excitatory pyramidal cells, which produces rhythmic fluctuations commonly referred to as \emph{oscilations}. The mechanisms that generate those oscillations is not yet completely understood, but it has been already identified that some `natural oscillations' provide evidence of activity being `processed' in certain regions of the brain at certain `frequencies'. Therefore, oscillatory behavior of human brain is often partitioned in bands (covering a wide range of frequencies decaying as $1/f$ in power): (\emph{i}) $\delta$-band (0.5-3Hz); (\emph{ii}) $\theta$-band (3.5-7Hz); (\emph{iii}) $\alpha$-band (8-13Hz); (\emph{iv}) $\beta$-band (14-30Hz); and (\emph{v}) $\gamma$-band (30-70Hz). Furthermore, there has been some evidence that activity in certain bands is associated with sensory registration, perception, movement and cognitive processes related to attention, learning and memory~\cite{bacsar2001gamma,engel2001dynamic,buzsaki2006rhythms}. Notwithstanding, such associations are often made using correlation and/or coherence techniques that only capture relationships between specific channels. But such methods are not able to assess the causality between signals that enables forecasting on the signal evolution, captured by a model-based representation as we propose to do hereafter. 

Different changes in the signals across different bands are also used to interpret the event-related potentials (ERPs) in the EEG signals, i.e., variations due to specific events -- see~\cite{wolpaw2012brain} for detailed analysis. In the context of sensory-motor data used in the current manuscript to validate the proposed methodology, sensorimotor rhythms (SMRs) are often considered. These represent oscillations that are recorded over the posterior frontal and anterior parietal areas of the brain, i.e., over the sensorimotor cortices (see Figure\,\ref{fig:EEGmontage}). SMRs occur mainly in the $\alpha$-band (for sensors located on the top of the motor cortices), and on beta and lower gamma for those on the sensorimotor cortices~\cite{pfurtscheller2001motor}. Consequently, these have been used as a \emph{default feature} for classification of motor-related execution and only the imagination of performing a motor task. Notwithstanding, the spatiotemporal modeling is simultaneously captured through direct state-space modeling that enables the system's understanding of the dynamics of the underlying process. In addition, it provides a new set of attributes that can be used to improve feature translation, i.e., classification.

\subsection{Spatiotemporal Fractional Model with Unknown Stimuli}
\label{ssec:sysMod}

A multitude of complex systems exhibit long-range (non-local) properties, interactions and/or dependencies (e.g., power-law decays in memories). Example of such systems includes Hamiltonian systems, where memory is the result of stickiness of trajectories in time to the islands of regular motion~\cite{4_mcmillen2008nonlinear}. Alternatively, it has been rigorously confirmed that viscoelastic properties are typical for a wide variety of biological entities like stem cells, liver, pancreas, heart valve, brain, muscles~\cite{2_kobayashi2012viscoelastic,3_wex2015experimental,4_mcmillen2008nonlinear,5_wang1993viscoelasticity,6_best1994characterization,7_doehring2005fractional,8_mace2011vivo,9_nicolle2012shear,10_grahovac2010modelling}, suggesting that memories of these systems obey the power law distributions. These dynamical systems can be characterized by the well-established mathematical theory of  fractional calculus~\cite{1_tarasov2011fractional}, and the corresponding systems could be described by fractional differential equations~\cite{11_bogdan2014heterogeneous,12_ghorbani2013cyber,13_xue2016spatio,14_xue2016minimum,15_xue2017constructing}.  However, it is until recently that fractional order system (FOS) starts to find its strong position in a wide spectrum of applications in different domains. This is due to the availability of computing and data acquisition methods to evaluate its efficacy in terms of capturing the underlying system states evolution. 

\begin{figure}[!h]
\centering
\includegraphics*[viewport=40 0 685 300, width = 3.35in, height = 1.4in]{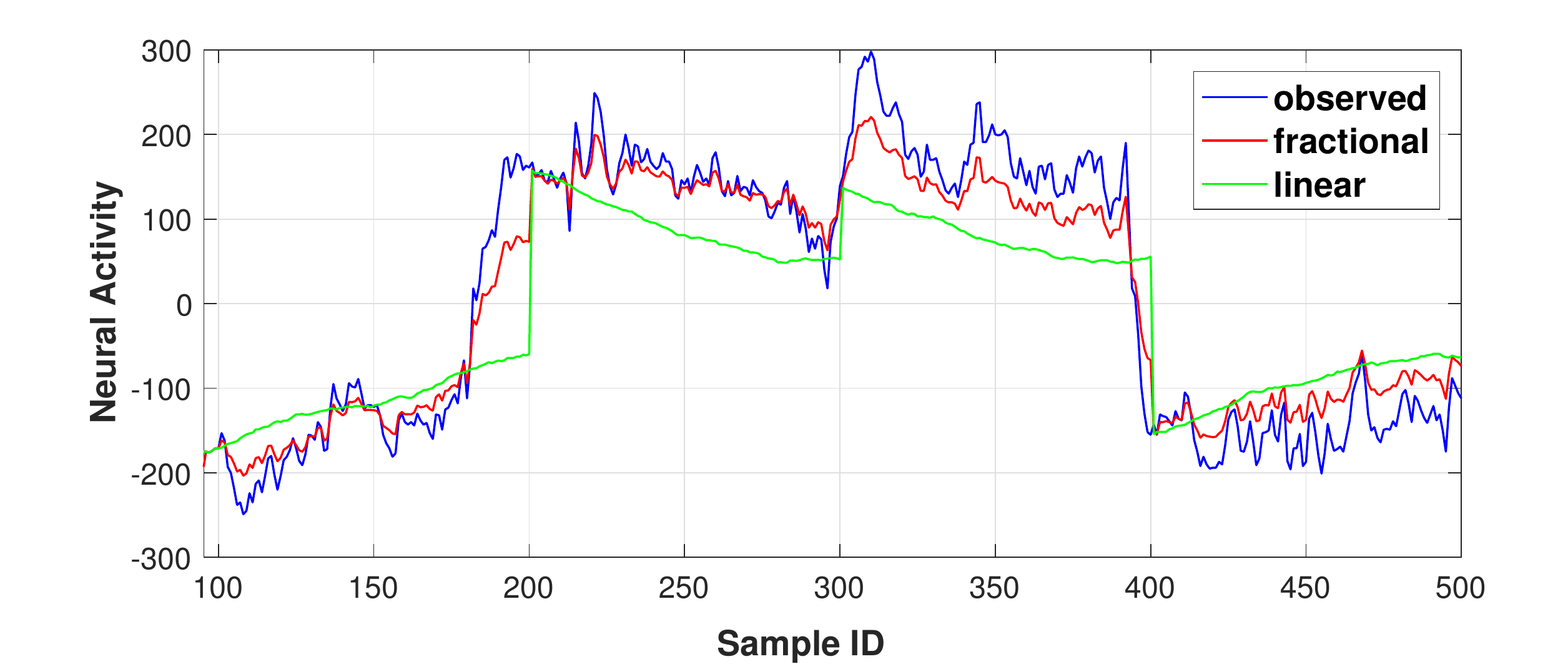}
\caption{Comparison of observed experimental EEG data with the prediction from fractional-order model and linear-time invariant model.}
\label{fig:fract_lin_comp}
\end{figure}

Subsequently, we consider a linear discrete time fractional-order dynamical model under unknown stimuli (i.e., inputs) described as follows:
\begin{eqnarray}
\Delta^{\alpha}x[k+1] &=& Ax[k] + Bu[k] \nonumber\\
y[k] &=& Cx[k], 
\label{eqn:fracLlinModel}
\end{eqnarray}
\noindent where $x\in\mathbb{R}^{n}$ is the state, $u \in \mathbb{R}^{p}$ is the unknown input and $y \in \mathbb{R}^{n}$ is the output vector. The differencing operator $\Delta$ is used as the discrete version of the derivative, for example $\Delta^{1}x[k] = x[k] - x[k-1]$. As evident, the difference order of $1$ has only one-step memory, and hence the classic linear-time invariant models are not able to answer the long-range memory property of several physiological signals as discussed before. On the other hand, the expansion of fractional-order derivative  in the discretized setting \cite{andrzej} for any $i$th state $(1\leq i\leq n)$ can be written as
\begin{equation}
\Delta^{\alpha_{i}}x_{i}[k] = \sum\limits_{j=0}^{k}\psi(\alpha_{i},j) x_{i}[k-j],
\label{eqn:fracExpan}
\end{equation}
\noindent where $\alpha_{i}$ is the fractional order corresponding to the $i$th state and $\psi(\alpha_{i},j) = \frac{\Gamma(j-\alpha_{i})}{\Gamma(-\alpha_{i})\Gamma(j+1)}$ with $\Gamma(.)$ denoting the gamma function. We can observe from (\ref{eqn:fracExpan}) that fractional-order derivate provide long-range memory by including all $x_{i}[k-j]$ terms. A quick comparison between the prediction accuracy of fractional-order derivative model and linear-time invariant model is shown in Figure\,\ref{fig:fract_lin_comp}. The fractional-order model can cope with sudden changes in the signals while the linear model cannot.

We can also describe the system by its matrices tuple $(\alpha,A, B, C)$ of appropriate dimensions. The coupling matrix $A$ represents the spatial coupling between the states across time while the input coupling matrix $B$ determines how inputs are affecting the states. In what follows, we assume that the input size is always strictly less than the size of state vector, i.e., $p < n$. 

Having defined the system model, the system identification, i.e., estimation of model parameters, from the given data is an important step. It becomes nontrivial when we have unknown inputs since one has to be able to differentiate which part of the evolution of the system is due to its intrinsic dynamics and what is due to the unknown inputs. Subsequently, the analysis part that we need to address is that of system identification from the data, as described next.

\subsection{Data driven system identification}
\label{ssec:sysIdenti}

The problem consists of estimating $\alpha$, $A$ and inputs $\{u[k]\}_{t}^{t+T-2}$ from the given limited observations $y[k]$, $k = [t, t + T -1]$, which due to the dedicated nature of sensing mechanism is same as $x[k]$ and under the assumption that the input matrix $B$ is known. The realization of $B$ can be application dependent and is computed separately using experimental data -- as we explore later in the case study, see Section~\ref{sec:experi}. For the simplicity of notation, let us denote $z[k] = \Delta^{\alpha}x[k+1]$ with $k$ chosen appropriately. The pre-factors in the summation in (\ref{eqn:fracExpan}) grows as $\psi(\alpha_{i},j) \sim \mathcal{O}(j^{-\alpha_{i}-1})$ and, therefore, for the purpose of computational ease we would be limiting the summation in (\ref{eqn:fracExpan}) to $J$ values, where $J>0$ is sufficiently large. Therefore, $z_{i}[k]$ can be written as
\begin{equation}
z_{i}[k] = \sum\limits_{j=0}^{J-1}\psi(\alpha_{i},j)x[k+1-j],
\label{eqn:zDefn}
\end{equation}
\noindent with the assumption that $x[k], u[k] = 0$ for $k\leq t-1$. Using the above introduced notations and the model definition in (\ref{eqn:fracLlinModel}), the given observations can be written as
\begin{equation}
z[k] = Ax[k] + Bu[k] + e[k],
\end{equation}
\noindent where $e \sim \mathcal{N}(0,\Sigma)$ is assumed to be  Gaussian noise independent across space and time. For simplicity we would assume that $\Sigma = \sigma^2 I$. Also, let us denote the system matrices as $A = [a_{1},a_{2},\hdots,a_{n}]^{T}$ and $B = [b_{1},b_{2},\hdots,b_{n}]^{T}$. The vertical concatenated states and inputs during an arbitrary window of time as  $X_{[t-1,t+T-2]} = [x[t-1],x[t],\hdots,x[t+T-2]]^{T}$, $U_{[t-1,t+T-2]} = [u[t-1],u[t],\hdots,u[t+T-2]]^{T}$ respectively, and for any $i$th state we have $Z_{i,[t-1,t+T-2]} = [z_{i}[t-1],z_{i}[t],\hdots,z_{i}[t+T-2]]^{T}$. For the sake of brevity we would be dropping the time horizon subscript from the above matrices as it is clear from the context.

Since the problem of joint estimation of the different parameters is highly nonlinear, we proceed as follows: (\emph{i}) we estimate the fractional order $\alpha$ using the wavelet technique described in \cite{flandrin}; and (\emph{ii}) with $\alpha$ known, the $z$ in equation (\ref{eqn:zDefn}) is computed under the additional assumption that the system matrix $B$ is known. Therefore, the problem now reduces to estimate $A$ and the inputs $\{u[k]\}_{t}^{t+T-2}$. Towards this goal, we exploit the algorithm similar to expectation-maximization (EM) \cite{McLachlam} from \cite{gaurav2017acc}. Briefly, the EM algorithm is used for maximum likelihood estimation (MLE) of parameters subject to hidden variables. Intuitively, in our case, in Algorithm~1, we estimate $A$ in the presence of hidden variables or \textit{unknown unknowns} $\{u[k]\}_{t}^{t+T-2}$. Therefore, the `E-step' is performed to average out the effects of unknown unknowns and obtain an estimate of $u$, where due to the diversity of solutions, we control the sparsity of the inputs using the parameter $\lambda'$. Subsequently, the `M-step' can then accomplish MLE estimation to obtain an estimate of $A$.

It was shown theoretically in \cite{gaurav2017acc} that the algorithm is convergent in the likelihood sense. It should also be noted that the EM algorithm can converge to saddle points as exemplified in \cite{McLachlam}. The Algorithm\,\ref{alg:EM_alg} being iterative is crucially dependent on the initial condition for the convergence. We will see in Section\,\ref{sec:experi} that the convergence is very fast making it suitable for online estimation of parameters.

%
%

\begin{algorithm}
\SetKw{KwInitialize}{Initialize}
\SetArgSty{normal}
\KwIn{ $x[k], k \in [t,t+T-1]$ and $B$}

\KwOut{$A$ and $\{u[k]\}_{t}^{t+T-2}$}

\KwInitialize{compute $\alpha$ using \cite{flandrin} and then $z[k]$}. For $l=0$, initialize $A^{(l)}$ as
\begin{equation*}
a_{i}^{(l)} = \text{arg}\min\limits_{a} \vert\vert Z_{i} - Xa\vert\vert_{2}^{2}
\end{equation*}
\Repeat{until converge}{

(i) \textbf{`E-step'}: For $k \in [t,t+T-2]$ obtain $u[k]$ as
\begin{equation*}
u[k] = \text{arg}\min\limits_{u}\vert\vert z[k] - A^{(l)}x[k] - Bu\vert\vert_{2}^{2} + \lambda'\vert\vert u\vert\vert_{1},
\end{equation*}
where $\lambda' = 2\sigma^{2}\lambda$;

(ii) \textbf{`M-step'}: \\ obtain $A^{(l+1)}= [a_{1}^{(l+1)},a_{2}^{(l+1)},\hdots,a_{n}^{(l+1)}]^{T}$ where
\begin{equation*}
a_{i}^{(l+1)} = \text{arg}\min\limits_{a} \vert\vert \tilde{Z}_{i} - Xa\vert\vert_{2}^{2},
\end{equation*}
\noindent and $\tilde{Z}_{i} = Z_{i} - Ub_{i}$\;
$l \leftarrow l + 1$\;
}
\caption{EM algorithm}
\label{alg:EM_alg}
\end{algorithm}
%

\subsection{Feature Translation (Classification)}
\label{ssec:featureClassi}

The unprocessed EEG signals coming from the sensors although carrying vital information may not be directly useful for making the predictions. However, by representing the signals in terms of parametric model $(\alpha, A)$ and the unknown signals as we did in the last section, we can gain better insights. The parameters of the model being representative of the original signal itself can be used to make a concise differentiation. 

The $A$ matrix represents how strong is the particular signal and how much it is affecting/being affected by the other signals that are considered together. While   performing or imagining particular motor tasks, certain regions of the brain gets more activated than others. Simultaneously, the inter-region activity also changes. Therefore, the columns of~$A$ which represent the coefficients of the strength of a signal affecting other signals can be used as a feature for classification of motor tasks. In this work, we will be considering the machine learning based classification techniques like logistic regression and Support Vector Machines~(SVM)~\cite{murphy}. The other classification techniques c The choice of kernels would vary from simple `linear' to radial basis function (RBF), i.e., $k(x_{i},x_{j}) = e^{-\gamma(x_{i}-x_{j})^2}$. The value of parameters of the classifier and possibly of the kernels are determined using the cross-validation. The range of parameters in the cross-validation are from $2^{-5},\hdots,2^{15}$ for $\gamma$ and $2^{-15},\hdots,2^{3}$ for $C = 1/\ \lambda$, both in the logarithmic scale, where $\lambda$ is the regularization parameter which appears in optimization cost of the classifiers \cite{murphy}.

\begin{figure}[b]
\centering
\begin{tikzpicture}[scale = 1.4]
\node[anchor=north west,inner sep=0] at (0,0) {\includegraphics*[viewport=10 40 580 630, width = 3.1in, height = 3in]{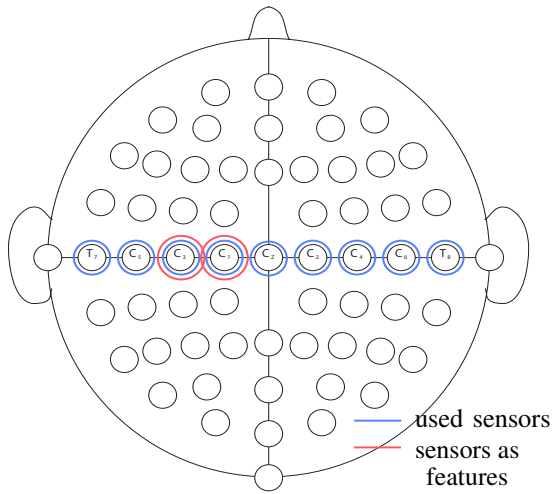}};
\node[anchor=north west] at (4.2,-3.85) {used sensors};
\node[anchor=north west] at (4.2,-4.20) {\Large$\substack{\text{sensors as}\\[0.4ex] \text{features}}$};
\end{tikzpicture}
\caption{A description of the sensor distribution for the measurement of EEG. The channel labels for the selected sensors are shown for dataset-I.}
\label{fig:usedSensors1}
\end{figure}

\section{Case Study}
\label{sec:experi}

We will now illustrate the usefulness of the \mbox{fractional-order} dynamic model with unknown inputs in the context of classification for BCI. We have considered two datasets from the BCI competition \cite{blankertz}. The datasets were selected on the priority of larger number of EEG channels and number of trials for training. The available data is split into the ratio of $60\%$ and $40\%$ for the purpose of training and testing, respectively.

\subsection{Dataset-I}
\label{ssec:ds_1}

We consider for validation the dataset labeled `dataset IVa' from BCI Competition-III~\cite{dornhege}. The recording was made using BrainAmp amplifiers and a $128$ channel electrode cap and out of which $118$ channels were used. The signals were band-pass filtered between $0.05$ and $200$ Hz and then digitized at $1000$ Hz. For the purpose of this study we have used the downsampled version at $100$ Hz. The dataset for subject ID `al' is considered, and it contains $280$ trials. The subject was provided a visual cue, and immediately after asked to imagine two motor tasks: (R) right hand, and (F) right foot.

\subsubsection{Sensor Selection and Modeling}
To avoid the curse-of-dimensionality, instead of considering 118 sensors available, which implies the use of $118\times 118$ dynamics entries for classification, only a subset of  $9$ sensors is considered. Specifically, only the sensors indicated in Figure\,\ref{fig:usedSensors1} are selected on the basis that only hand and feet movements need to be predicted, and only a $9\times 9$ dynamics matrix and $9$ fractional order coefficients are required for modeling the fractional order system. Besides, these sensors are selected because they are close to the region of the brain known to be associated with motor actions. 

\begin{figure}[!h]
\centering
\includegraphics*[viewport=50 0 650 230, width = 3.35in, height = 1.5in]{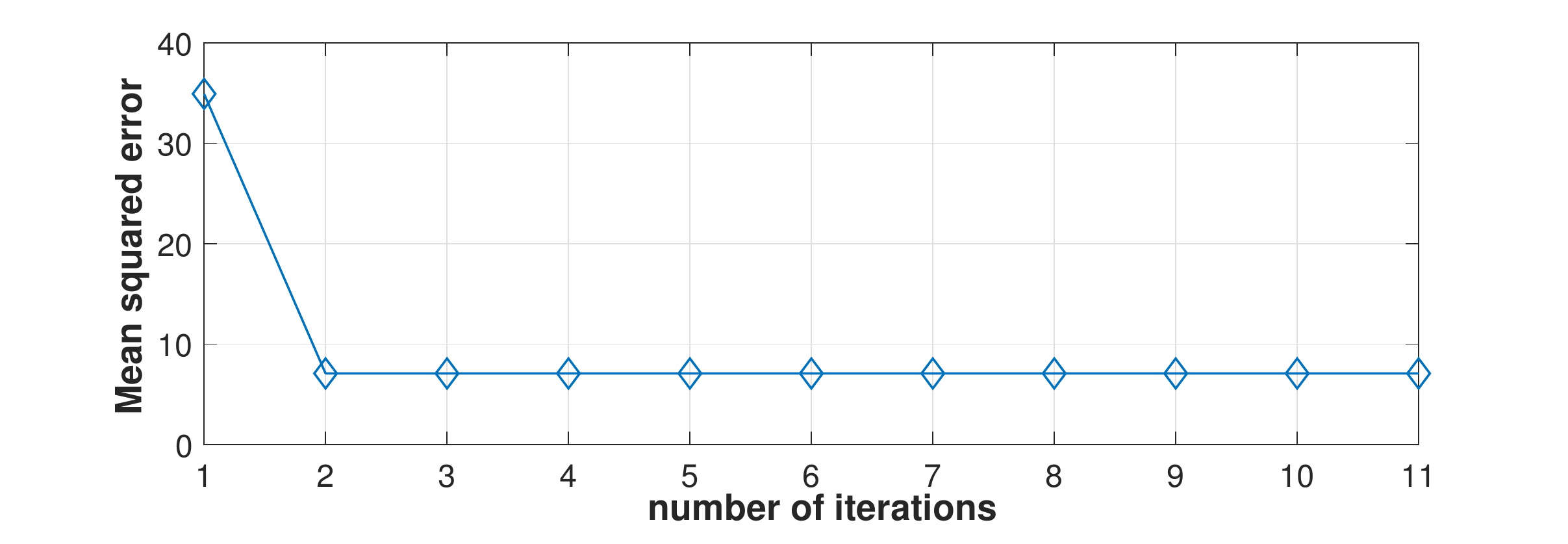}
\caption{Mean squared error of Algorithm\,\ref{alg:EM_alg} as function of number of iterations for dataset-I.}
\label{fig:mse1}
\end{figure}

\subsubsection{System Identification and Validation}
\begin{figure}[!h]
\centering
\begin{subfigure}[t]{0.5\textwidth}
\includegraphics*[viewport=0 230 600 490, width = 3.35in, height = 1.5in]{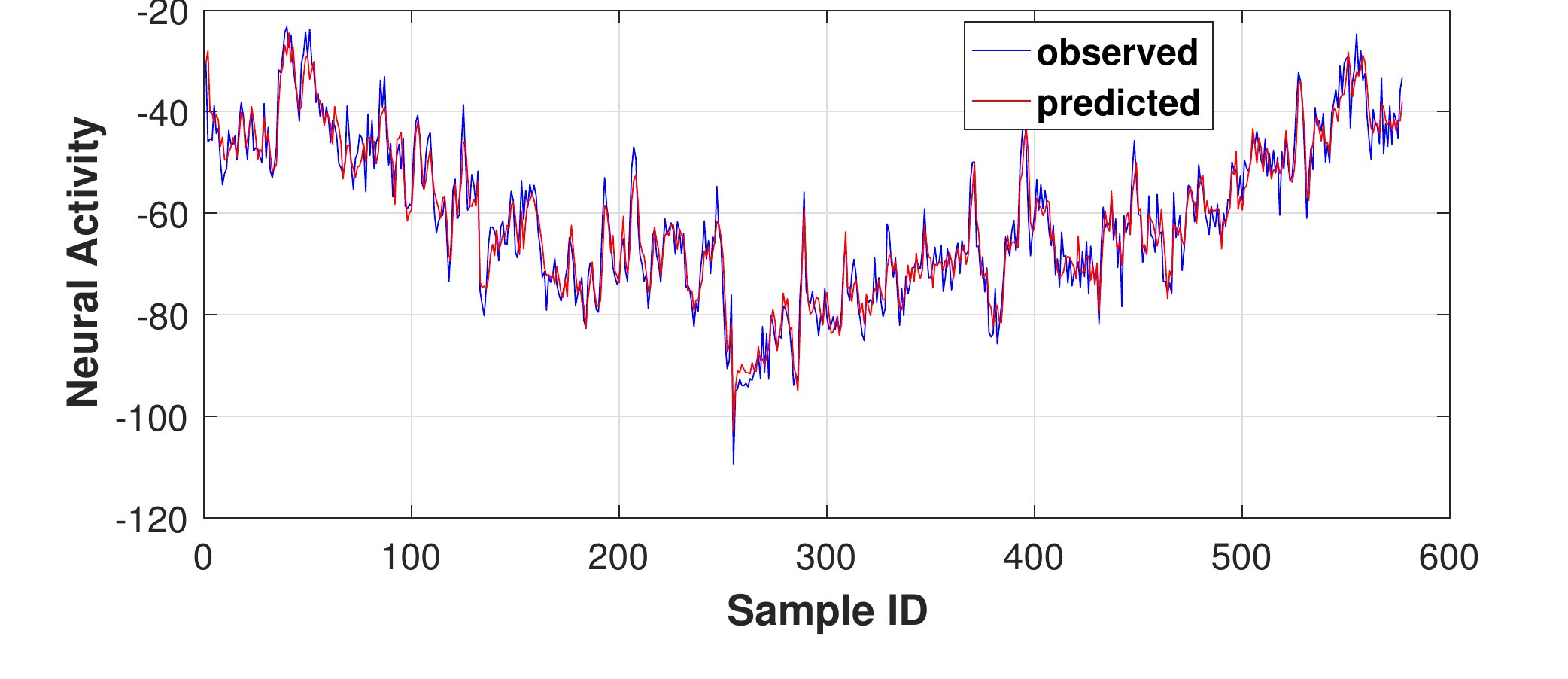}
\caption{}
\label{sfig:oneStep1}
\end{subfigure}

\begin{subfigure}[t]{0.5\textwidth}
\includegraphics*[viewport=0 230 600 490, width = 3.35in, height = 1.5in]{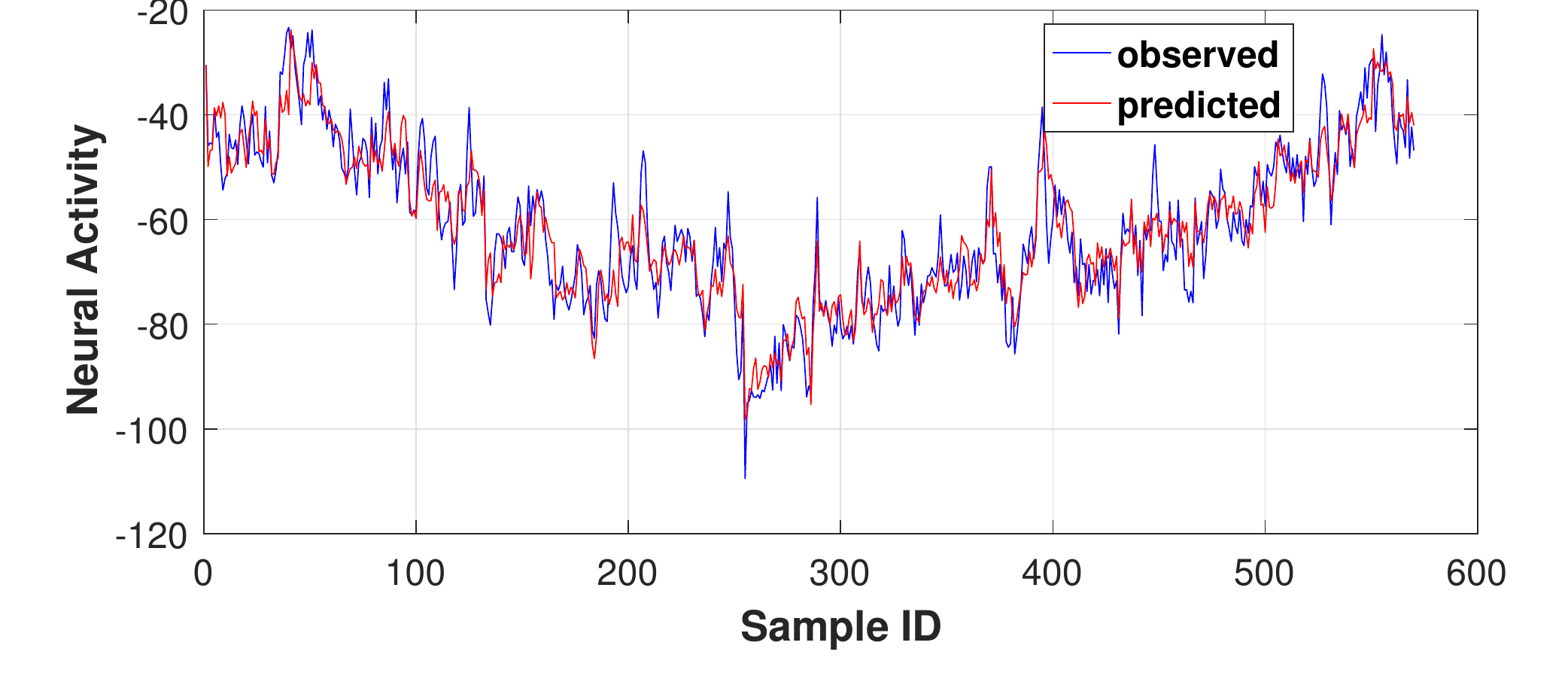}
\caption{}
\label{sfig:fiveStep1}
\end{subfigure}
\caption{Comparison of predicted EEG state for the channel $T_{7}$ using fractional-order dynamical model with unknown inputs. The one step and five step predictions are shown in (a) and (b) respectively.}
\label{fig:pred1}
\end{figure}

The model parameters $(\alpha,A)$ and the unknown inputs are estimated by using the Algorithm\,\ref{alg:EM_alg}. As mentioned before, the performance of the algorithm being iterative is dependent on the choice of the initial conditions. For the current case, we have observed that the algorithm converges very fast, and even a single iteration is enough. The convergence of mean squared error in the M-step of Algorithm\,\ref{alg:EM_alg} for one sample from dataset-I is shown in Figure\,\ref{fig:mse1}. This shows that the choice of initial conditions are fairly good. The one step and five step prediction of the estimated model is shown in Figure\,\ref{fig:pred1}. It is evident that the predicted values for one step very closely follow the actual values. There are some differences between the actual and predicted values for five step prediction.

\subsubsection{Discussion of the results} 
The most popular features used in the motor-imagery based BCI classification relies on exploiting the spectral information. The features used are the band-power which quantifies the energy of the EEG signals in certain spectrum bands \cite{bashashati,brodu1,herman}. The motor cortex of the brain is known to be affecting the energy in the bands namely, $\alpha$ and $\beta$ as discussed in Section\,\ref{sec:probForm}. While it happens that unwanted signal energy is captured in these bands as well while performing the experiments, for example neck movement, other muscle activities etc. The filtering of these so called `unwanted' components from the original signal is a challenging task using the spectral techniques as they often share the same band. 

\begin{figure}
\centering
\includegraphics*[viewport=100 90 760 640, width = 3.35in, height = 3in]{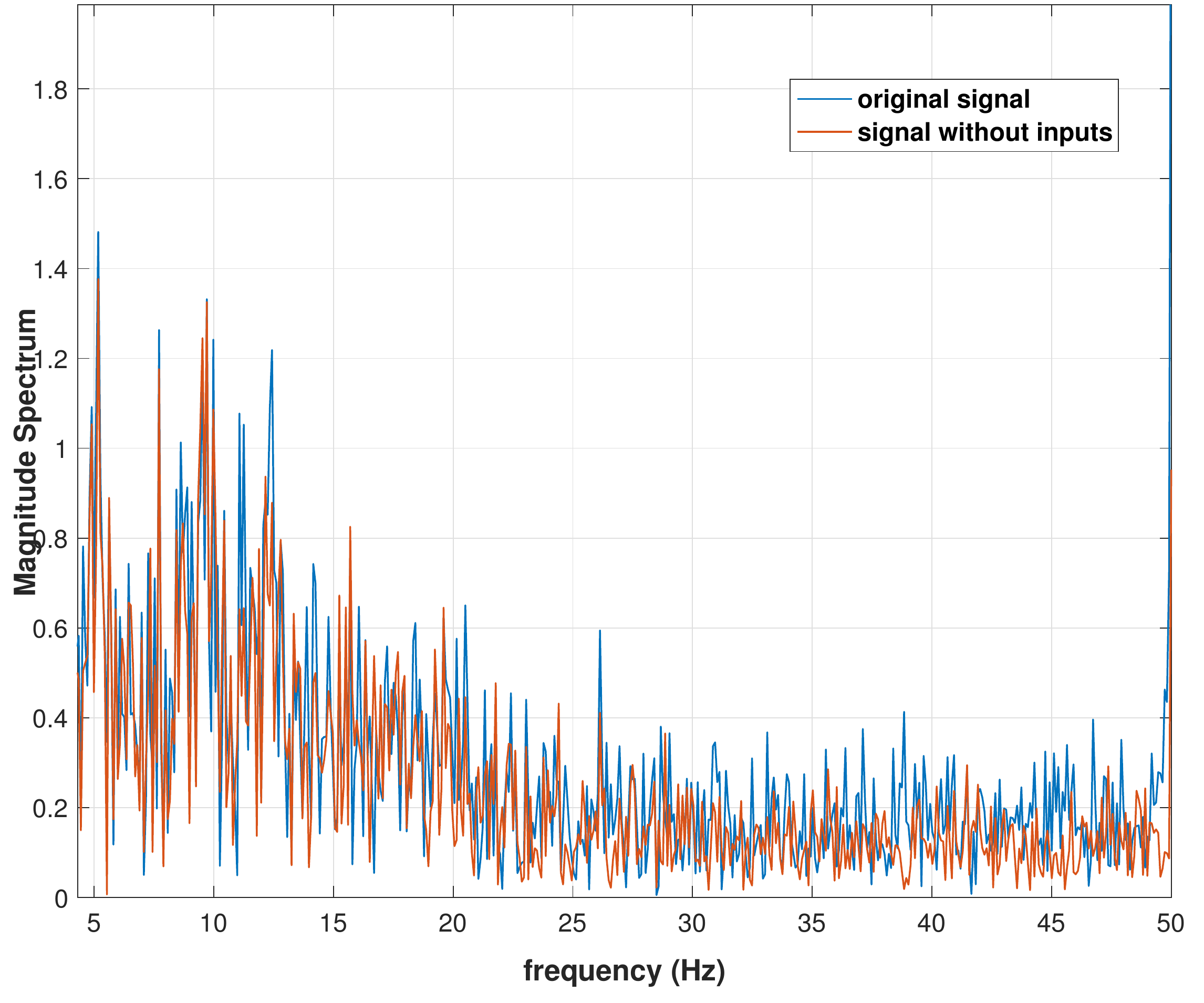}
\caption{Magnitude spectrum of the signal recorded by channel $T_{7}$ with and without unknown inputs.}
\label{fig:spect1}
\end{figure}

\begin{figure}[!h]
\centering
\includegraphics*[viewport=20 30 650 500, width = 2.7in, height = 2.3in]{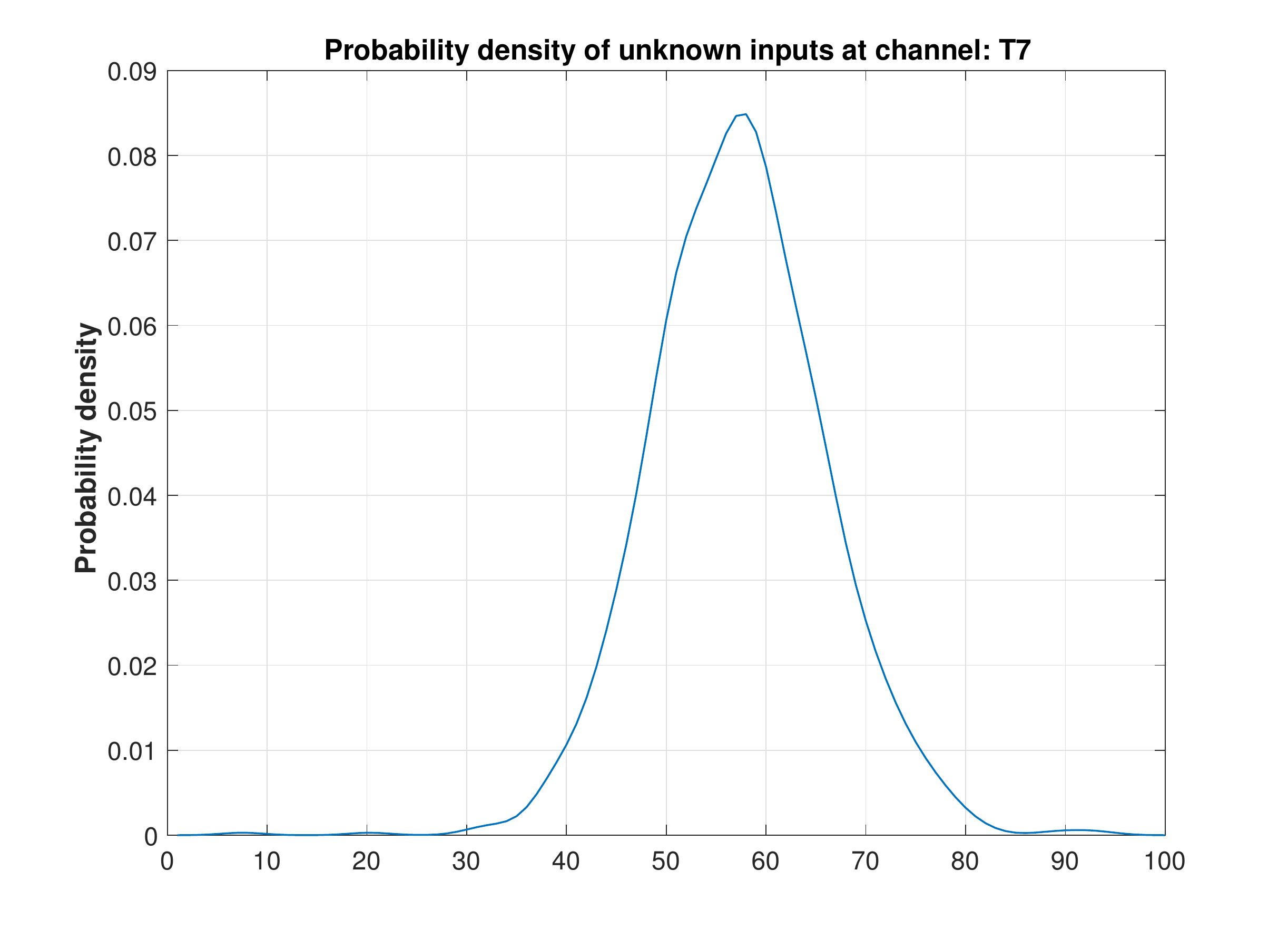}
\caption{Probability density function of the unknown inputs estimated from the signal recorded by channel $T_{7}$.}
\label{fig:inpPdf1}
\end{figure}

We used a different approach to deal with these unknown unknowns in Section\,\ref{sec:probForm}. The magnitude spectrum of the original EEG signal and on removing the estimated unknown inputs is shown in Figure\,\ref{fig:spect1}. It should be observed that the original signal and the signal upon removing the unknown inputs have significant energy in the $\alpha$ and $\beta$ bands. The unknown inputs behave similar to the white noise which is evident from their Gaussian probability distribution (PDF) as shown in Figure\,\ref{fig:inpPdf1}. The inputs are not mean zero but their PDF is centered around a mean value of approximately 58.

\begin{figure}
\centering
\begin{tikzpicture}[scale = 1.4]

\node[anchor=north west,inner sep=0] at (0,0) {\includegraphics*[viewport=30 25 470 430, width = 1.5in, height = 1.7in]{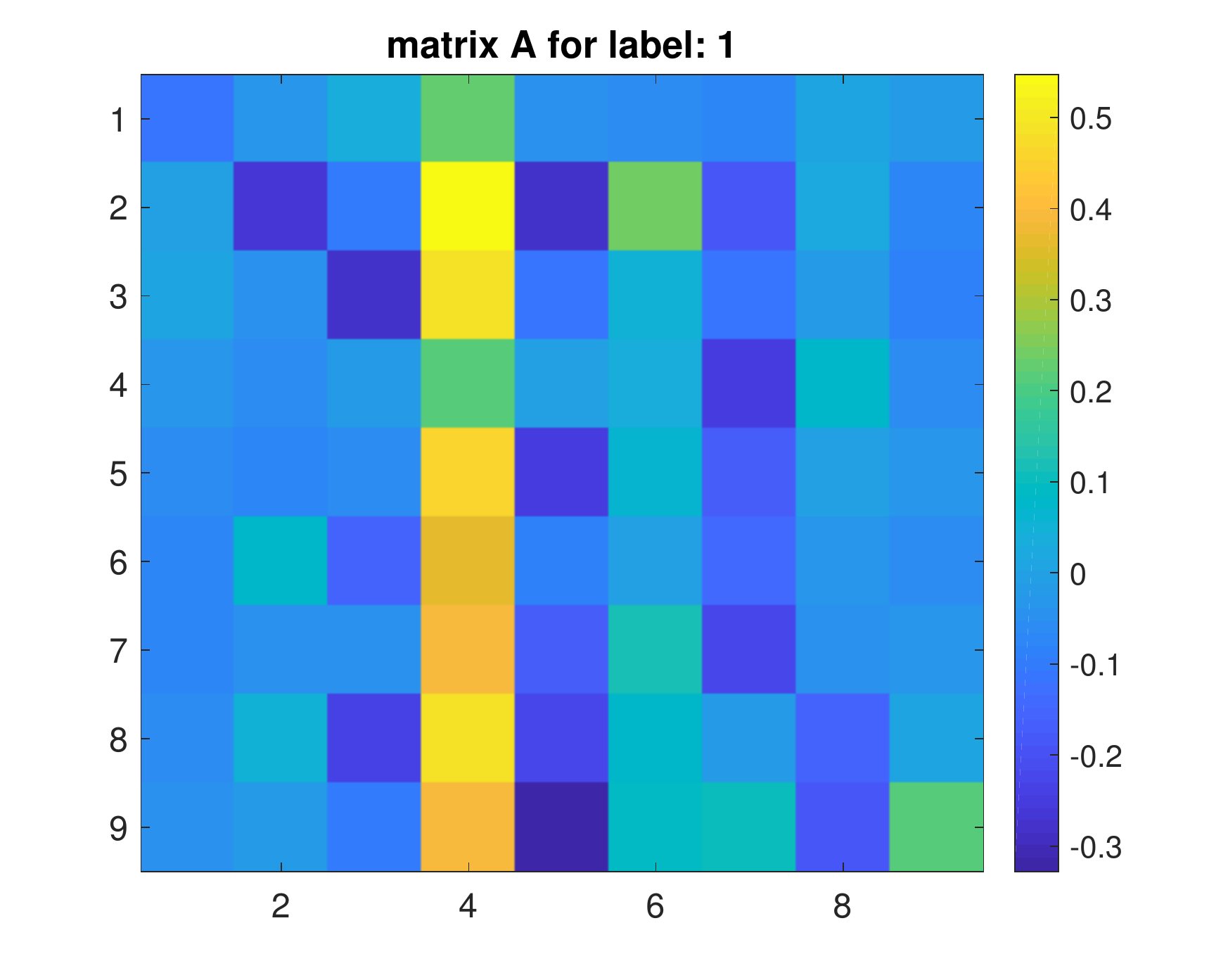}};

\node[anchor=north west,inner sep=0] at (2.8,0) {\includegraphics*[viewport=30 25 470 430, width = 1.5in, height = 1.7in]{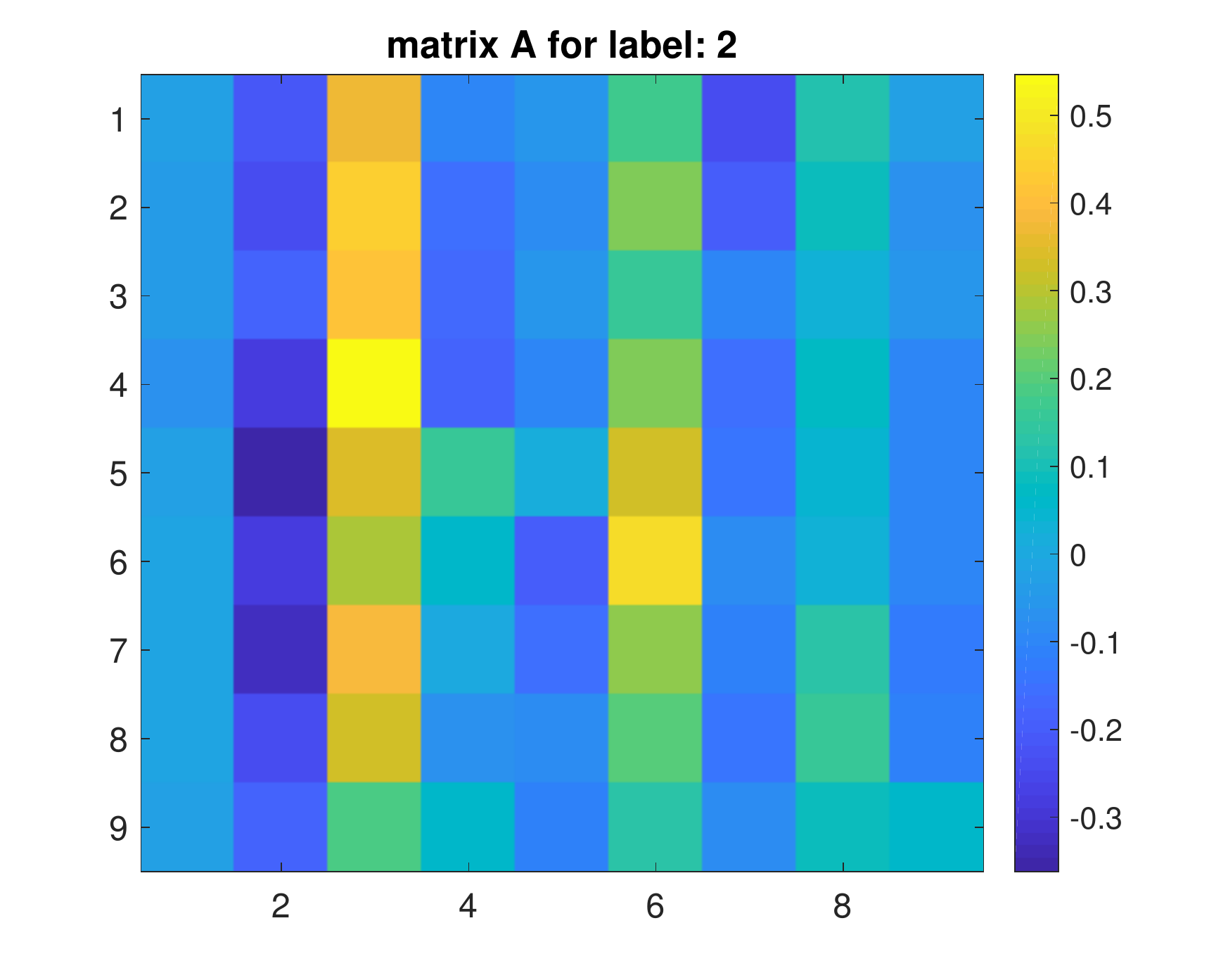}};

\draw [anchor=north, color=red, line width=1pt] (0.65,-3.15) rectangle (1.1,-.43);
\draw [anchor=north, color=red, line width=1pt] (0.65+2.8,-3.15) rectangle (1.1+2.8,-.43);

\end{tikzpicture}
\caption{Estimated $A$ matrix of size $9\times 9$ for the dataset-I with marked columns corresponding to the sensor index $3$ and $4$ used for classification.}
\label{fig:matA1}
\end{figure}

The model parameters $(\alpha, A)$ are jointly estimated with the unknown inputs using Algorithm\,\ref{alg:EM_alg}, therefore the effect of the inputs is inherently taken care of in the parameters. The structure of matrix $A$ for two different labels is shown in Figure\,\ref{fig:matA1}. We have used the sensors $C_{3}$ and $C_{1}$ which are indexed as $3$ and $4$, respectively in Figure\,\ref{fig:matA1}. It is apparent from Figure\,\ref{fig:matA1} that the columns corresponding to these sensors have different activity and hence deem to be fair candidates for the features to be used in classification. Therefore, the total number of features are $2\times 9=18$.

\subsection{Dataset-II}
\label{ssec:ds_2}

A $118$ channel EEG data from BCI Competition-III, labeled as `dataset IVb' is taken \cite{dornhege}. The data acquisition technique and sampling frequencies are same as in dataset of the previous subsection. The total number of labeled trials are $210$. The subjects upon provided visual cues were asked to imagine two motor tasks, namely (L) left hand and (F) right foot.

\subsubsection{Sensor Selection and Modeling}
\begin{figure}[t]
\centering
\begin{tikzpicture}[scale = 1.4]
\node[anchor=north west,inner sep=0] at (0,0) {\includegraphics*[viewport=10 40 580 630, width = 3.1in, height = 3in]{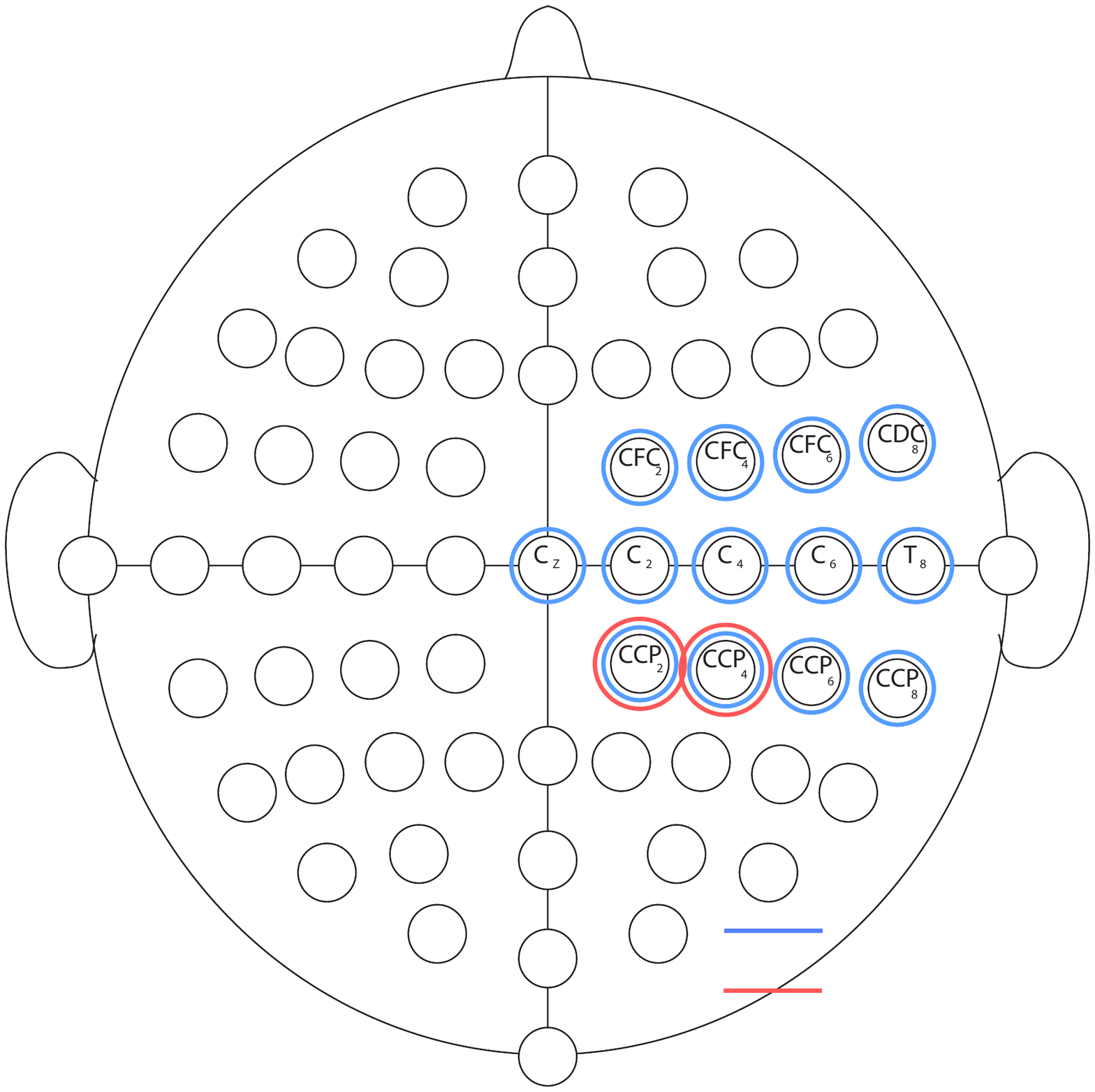}};
\node[anchor=north west] at (4.2,-3.85) {used sensors};
\node[anchor=north west] at (4.2,-4.20) {\Large$\substack{\text{sensors as}\\[0.4ex] \text{features}}$};
\end{tikzpicture}
\caption{A description of the sensor distribution for the measurement of EEG. The channel labels for the selected sensors are shown for dataset-II.}
\label{fig:usedSensors2}
\end{figure}

Due to the small number of training examples, we have again resorted to select the subset of sensors for the model estimation as we did for the dataset-I in the previous section. Since the motor tasks were left hand and feet, therefore we have selected the sensors in the right half of the brain and close to the region which is known to be associated with hand and feet movements as shown in Figure\,\ref{fig:usedSensors2}. We will see in the final part of this section that selecting sensors based on such analogy helps not only in reducing the number of features, but also to gain better and meaningful results.

\begin{figure}[!h]
\includegraphics*[viewport=45 0 650 230, width = 3.35in, height = 1.5in]{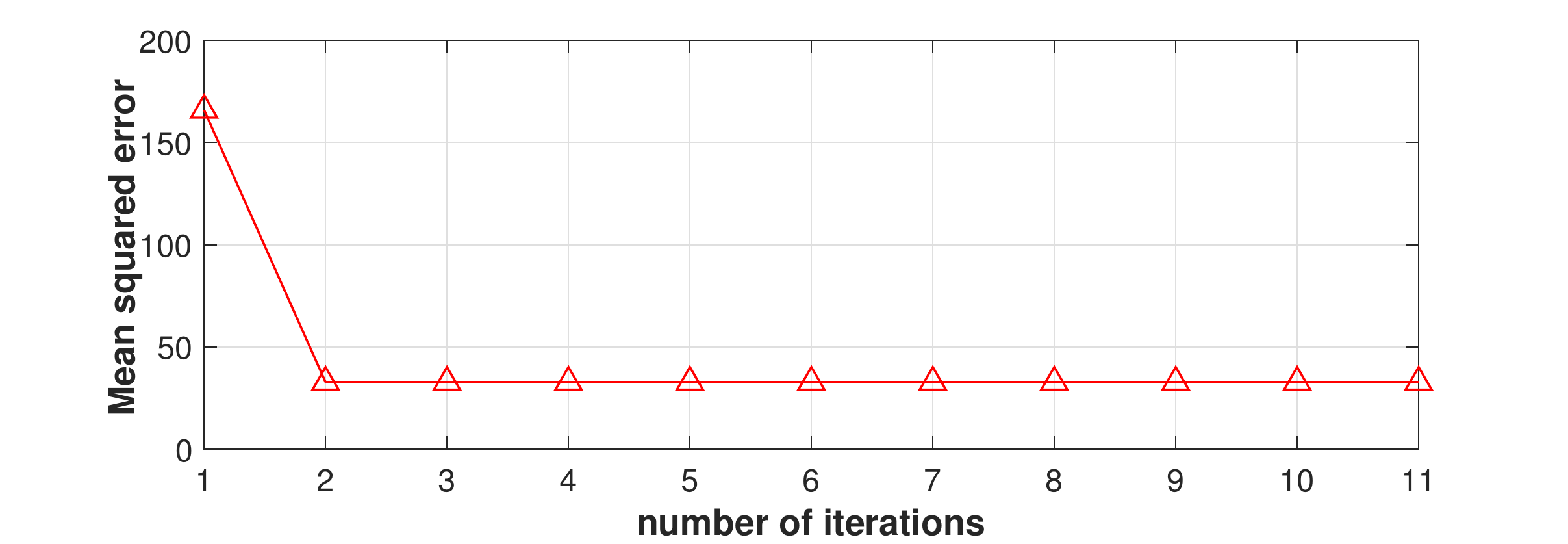}
\caption{Mean squared error of Algorithm\,\ref{alg:EM_alg} as function of number of iterations for dataset-II.}
\label{fig:mse2}
\end{figure}

\begin{figure}[!h]
\centering
\begin{subfigure}[t]{0.5\textwidth}
\includegraphics*[viewport=0 230 600 490, width = 3.35in, height = 1.5in]{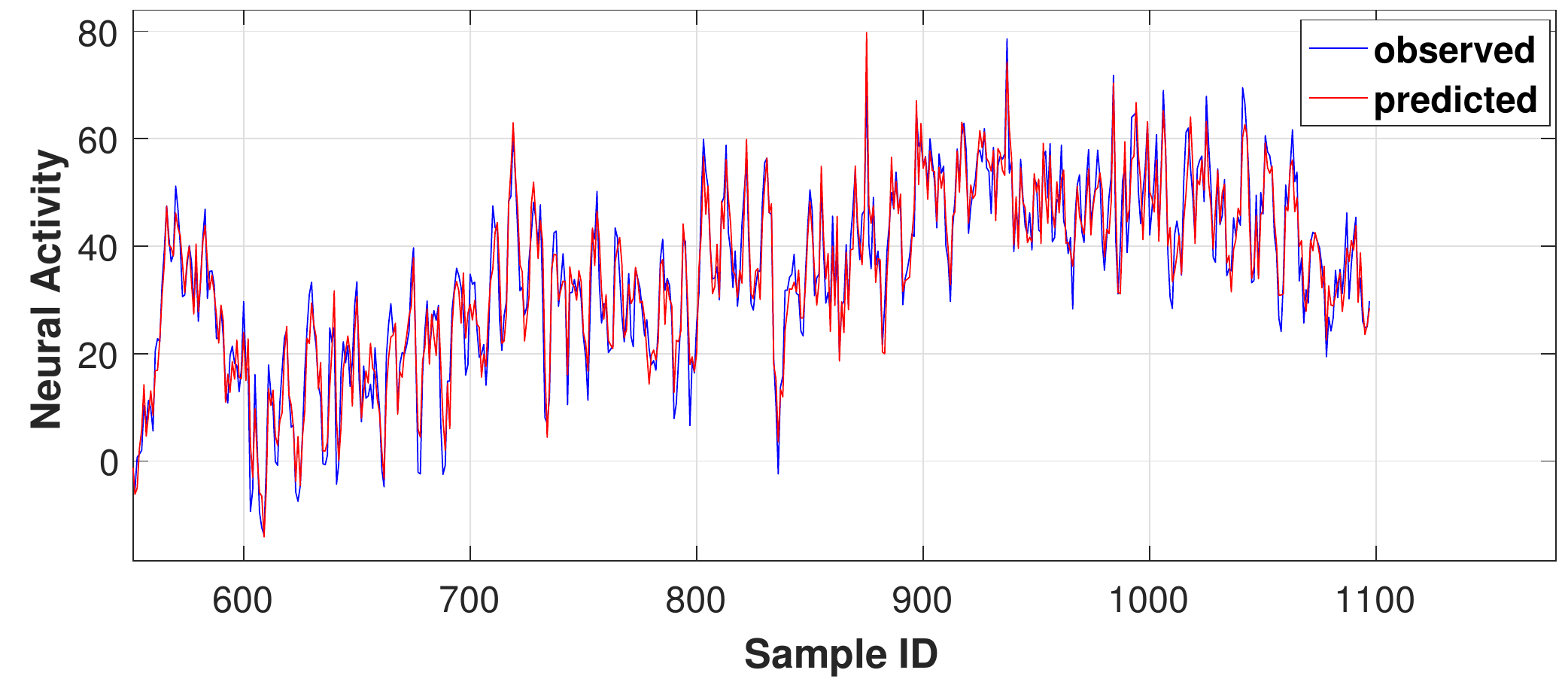}
\caption{}
\label{sfig:oneStep2}
\end{subfigure}

\begin{subfigure}[t]{0.5\textwidth}
\includegraphics*[viewport=0 230 600 490, width = 3.35in, height = 1.5in]{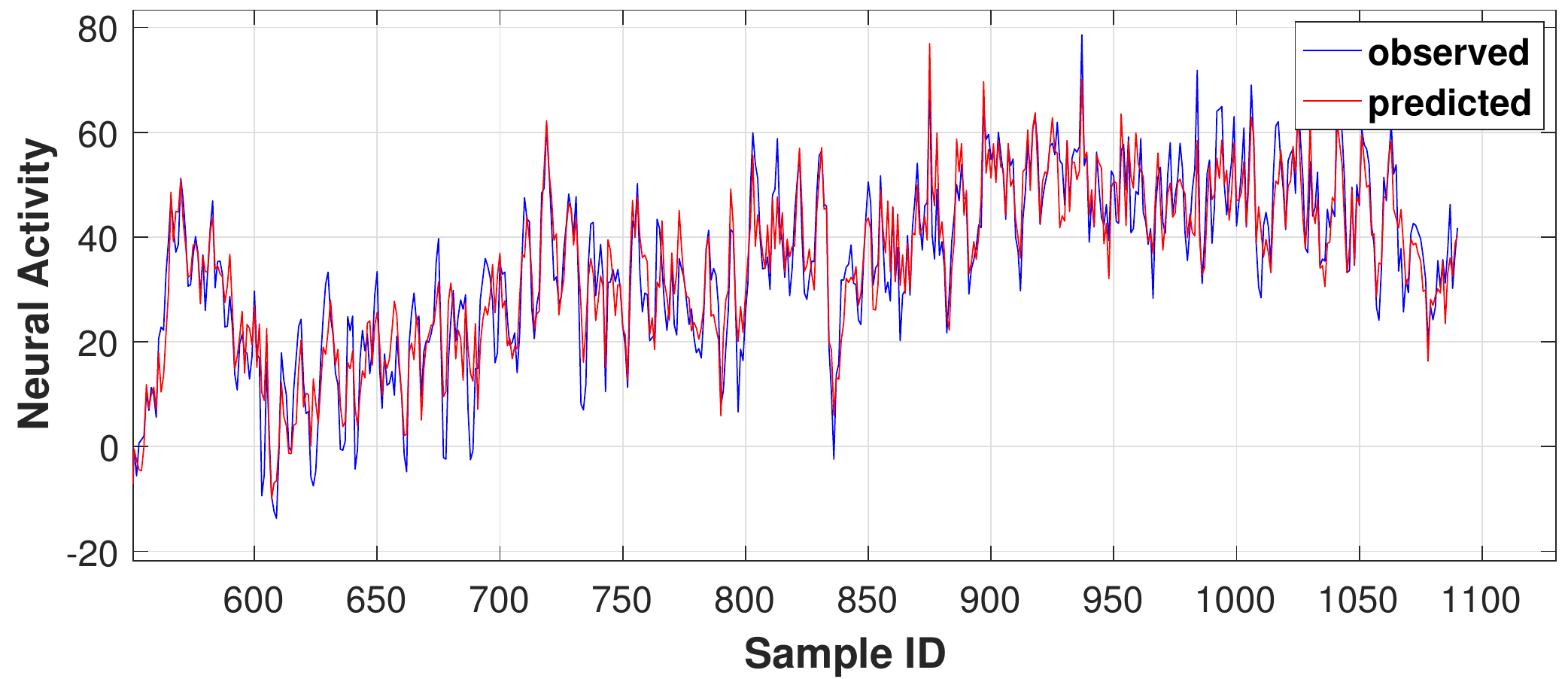}
\caption{}
\label{sfig:fiveStep2}
\end{subfigure}
\caption{Comparison of predicted EEG state for the channel $CFC_{2}$ using fractional-order dynamical model with unknown inputs. The one step and five step predictions are shown in (a) and (b) respectively.}
\label{fig:pred2}
\end{figure}

\subsubsection{System Identification and Validation}
After performing the estimation of the model $(\alpha, A)$ and the unknown inputs using the subset of sensors, we can see the similar performance of the model on dataset-II as was in dataset-I. The convergence of mean squared error in the M-step of Algorithm\,\ref{alg:EM_alg} for one sample from dataset-II is shown in Figure\,\ref{fig:mse2}. The one step and five step predictions are shown in Figure\,\ref{fig:pred2}. The model prediction follows closely the original signal.
\begin{figure}
\centering
\includegraphics*[viewport=100 100 760 640, width = 3.35in, height = 3in]{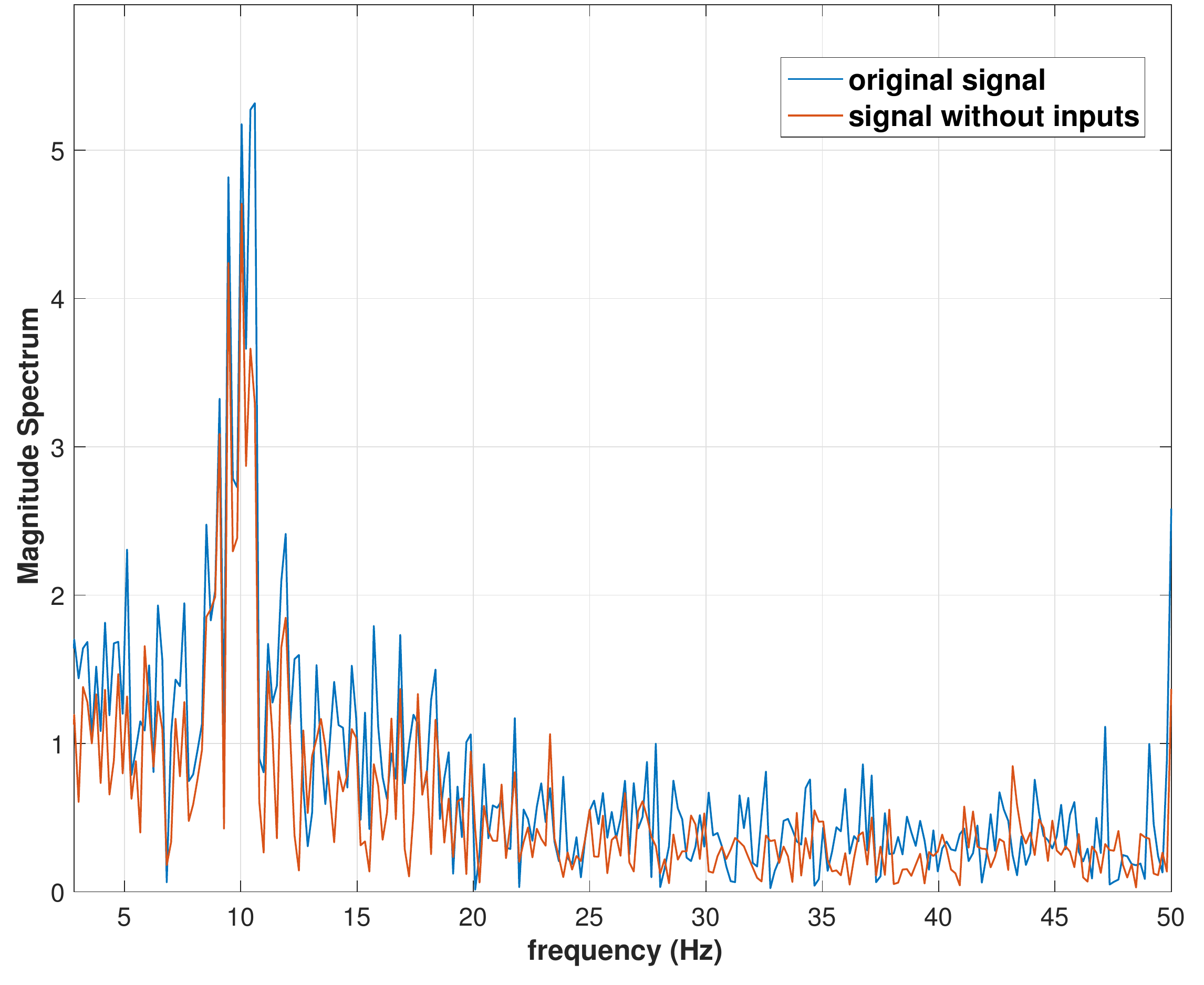}
\caption{Magnitude spectrum of the signal recorded by channel $CFC_{2}$ with and without unknown inputs.}
\label{fig:spect2}
\end{figure}

\begin{figure}
\centering
\includegraphics*[viewport=20 30 650 500, width = 2.7in, height = 2.3in]{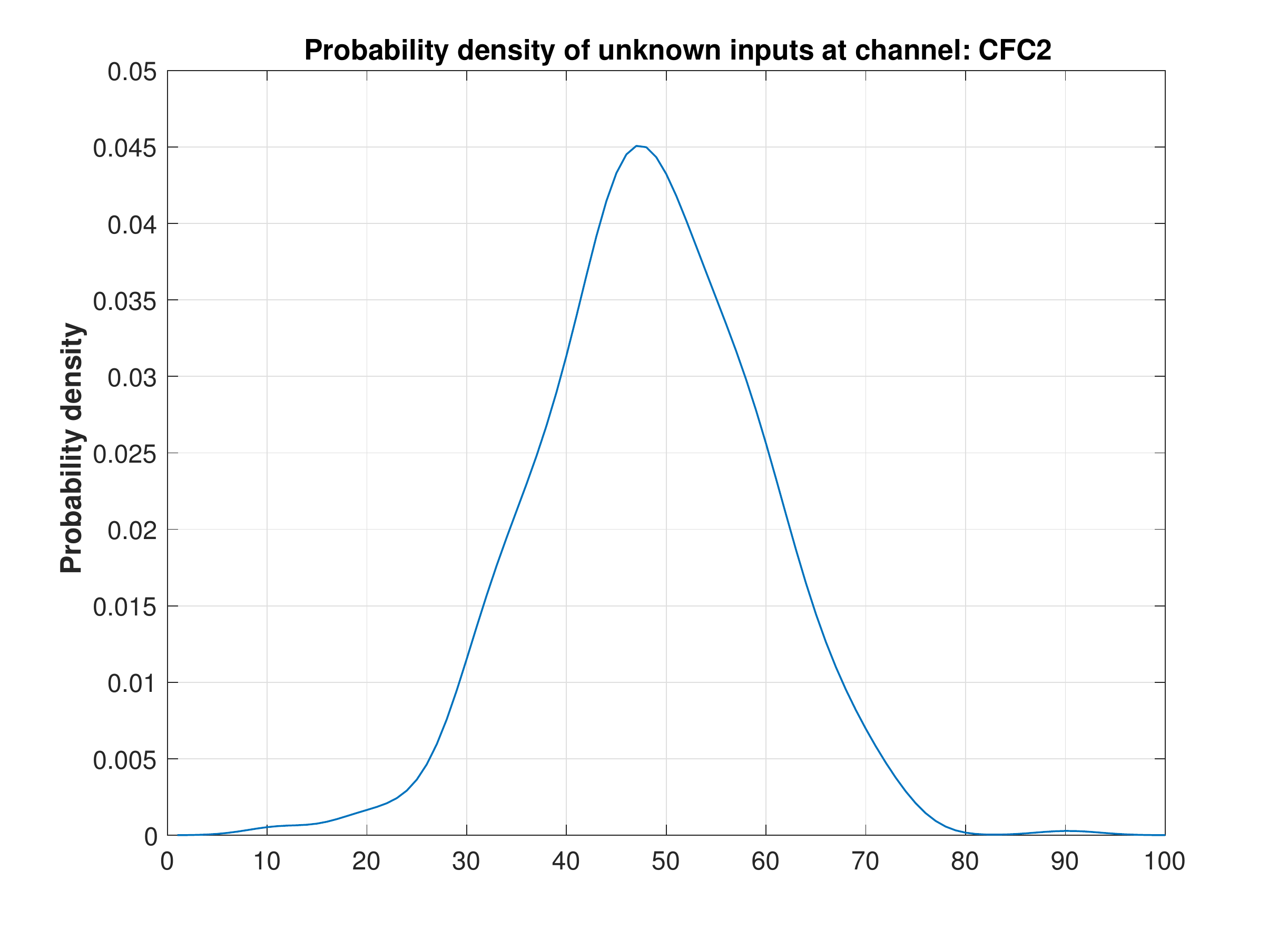}
\caption{Probability density function of the unknown inputs estimated from the signal recorded by channel $CFC_{2}$.}
\label{fig:inpPdf2}
\end{figure}

\begin{figure}[t]
\centering
\begin{tikzpicture}[scale = 1.4]

\node[anchor=north west,inner sep=0] at (0,0) {\includegraphics*[viewport=30 25 470 430, width = 1.5in, height = 1.7in]{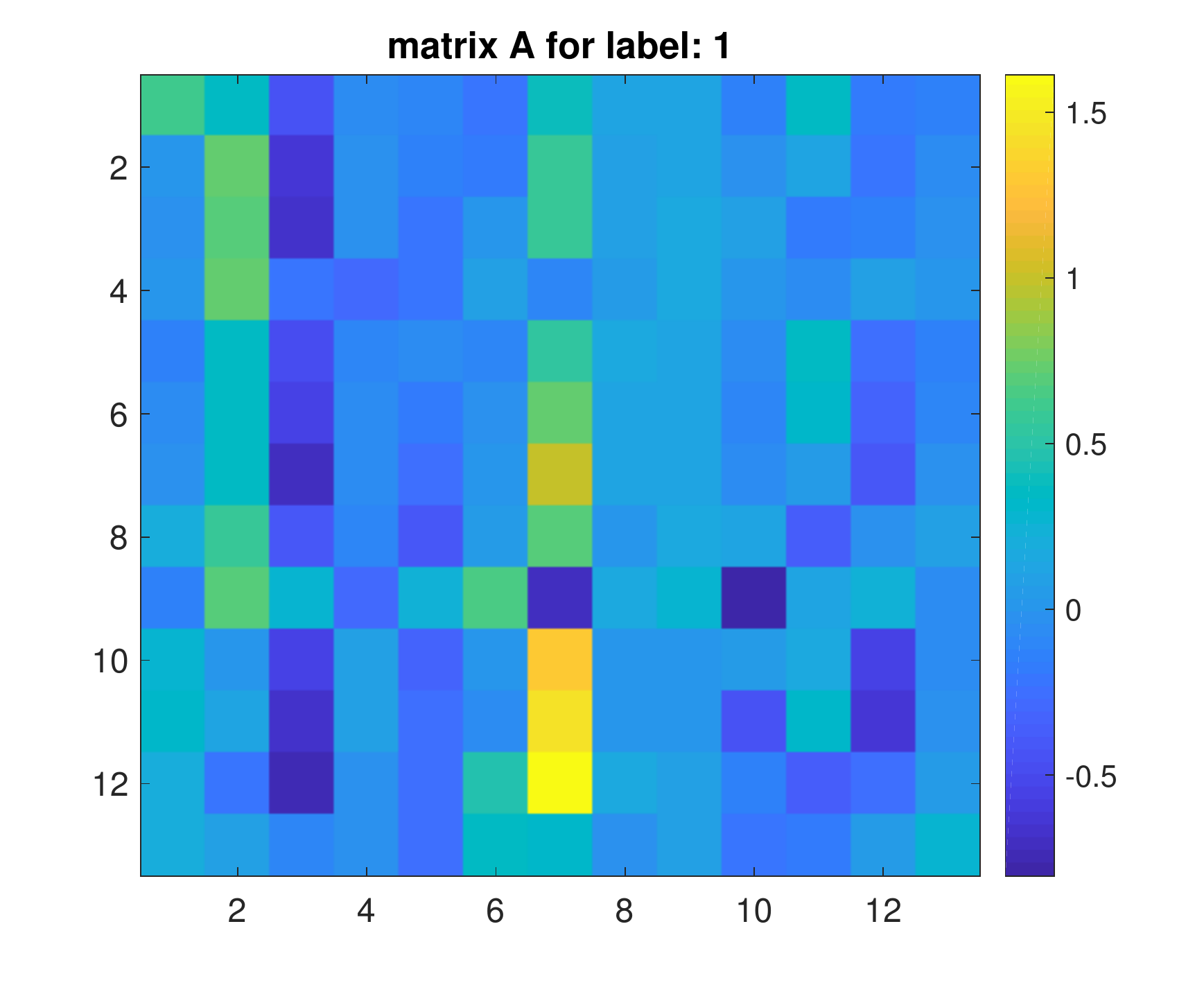}};

\node[anchor=north west,inner sep=0] at (2.8,0) {\includegraphics*[viewport=30 25 470 430, width = 1.5in, height = 1.7in]{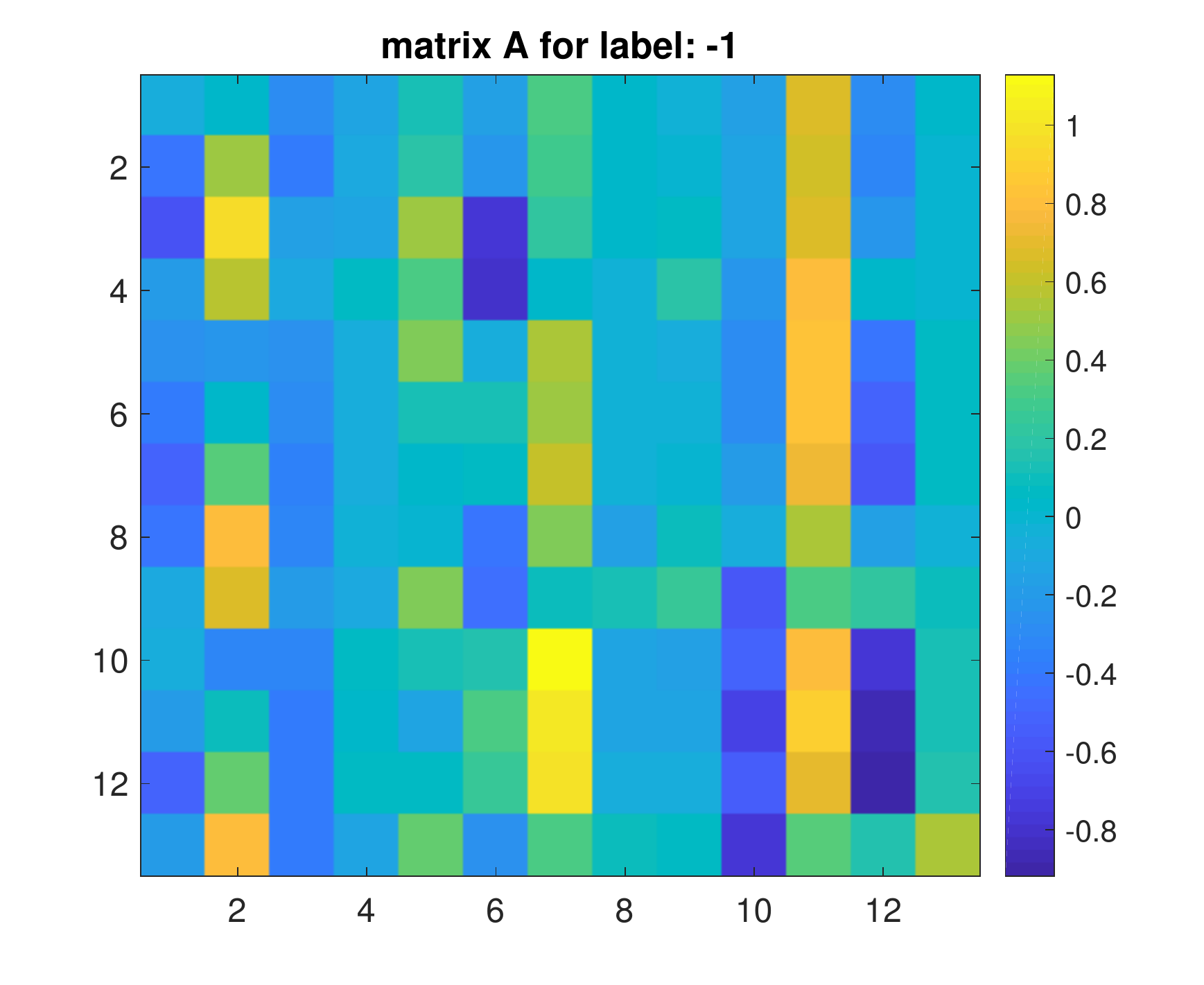}};

\draw [anchor=north, color=red, line width=1pt] (1.65,-3.15) rectangle (2.05,-.38);
\draw [anchor=north, color=red, line width=1pt] (1.65+2.8,-3.15) rectangle (2.05+2.8,-.38);

\end{tikzpicture}
\caption{Estimated $A$ matrix of size $13\times 13$ for the dataset-II with marked columns corresponding to the sensor index $10$ and $11$ used for classification.}
\label{fig:matA2}
\end{figure}

\subsubsection{Discussion of the results} 
The spectrum of the original EEG signal at channel $CFC_{2}$ and its version with unknown inputs removed are shown in Figure\,\ref{fig:spect2}. The spectrum shows peaks in the $\alpha$ and $\beta$ bands. We witness a similar observation as before that both of the signals share the same band and hence making it difficult to remove the effects of the unwanted inputs. The unknown inputs resembles that of white noise and the PDF is close to Gaussian distribution with mean centered at around 48.

The estimated $A$ matrix from Algorithm\,\ref{alg:EM_alg} is shown in Figure\,\ref{fig:matA2} for two different labels. Out of all $13$ sensors, the sensors $CCP_{2}$ and $CCP_{4}$ which are indexed as $10$ and $11$ in the matrix have striking different activity. The columns corresponding to these two sensors seem good choice for being the features for classification. Therefore, the total number of features are $2\times 13 = 26$ for this dataset. Next, we discuss the classification accuracy for both the datasets.

\begin{figure}[!h]
\centering
\begin{subfigure}[t]{0.5\textwidth}
\includegraphics*[viewport=0 0 685 350, width = 3.35in, height = 1.8in]{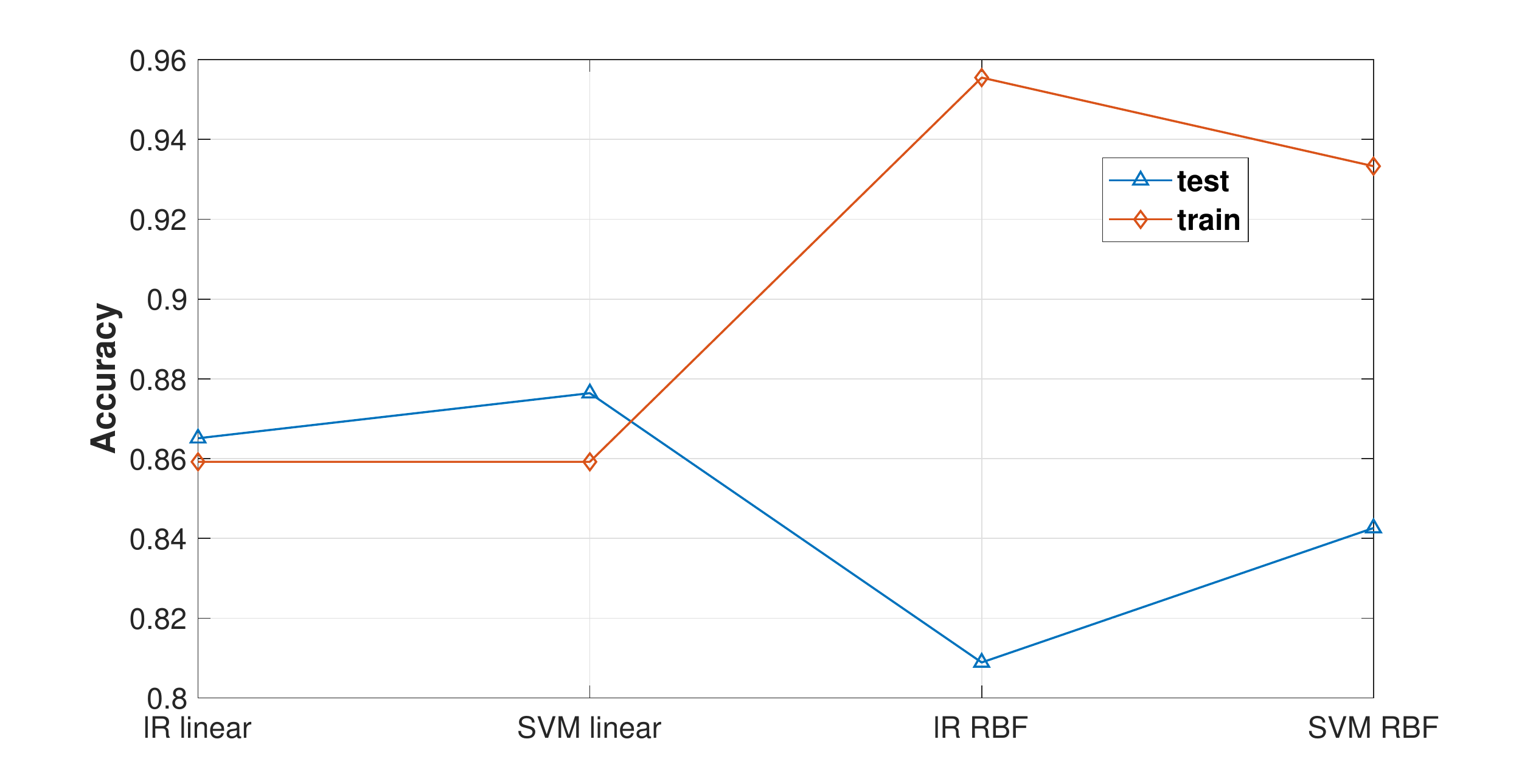}
\caption{}
\label{sfig:acc1}
\end{subfigure}

\begin{subfigure}[t]{0.5\textwidth}
\includegraphics*[viewport=0 0 685 350, width = 3.35in, height = 1.8in]{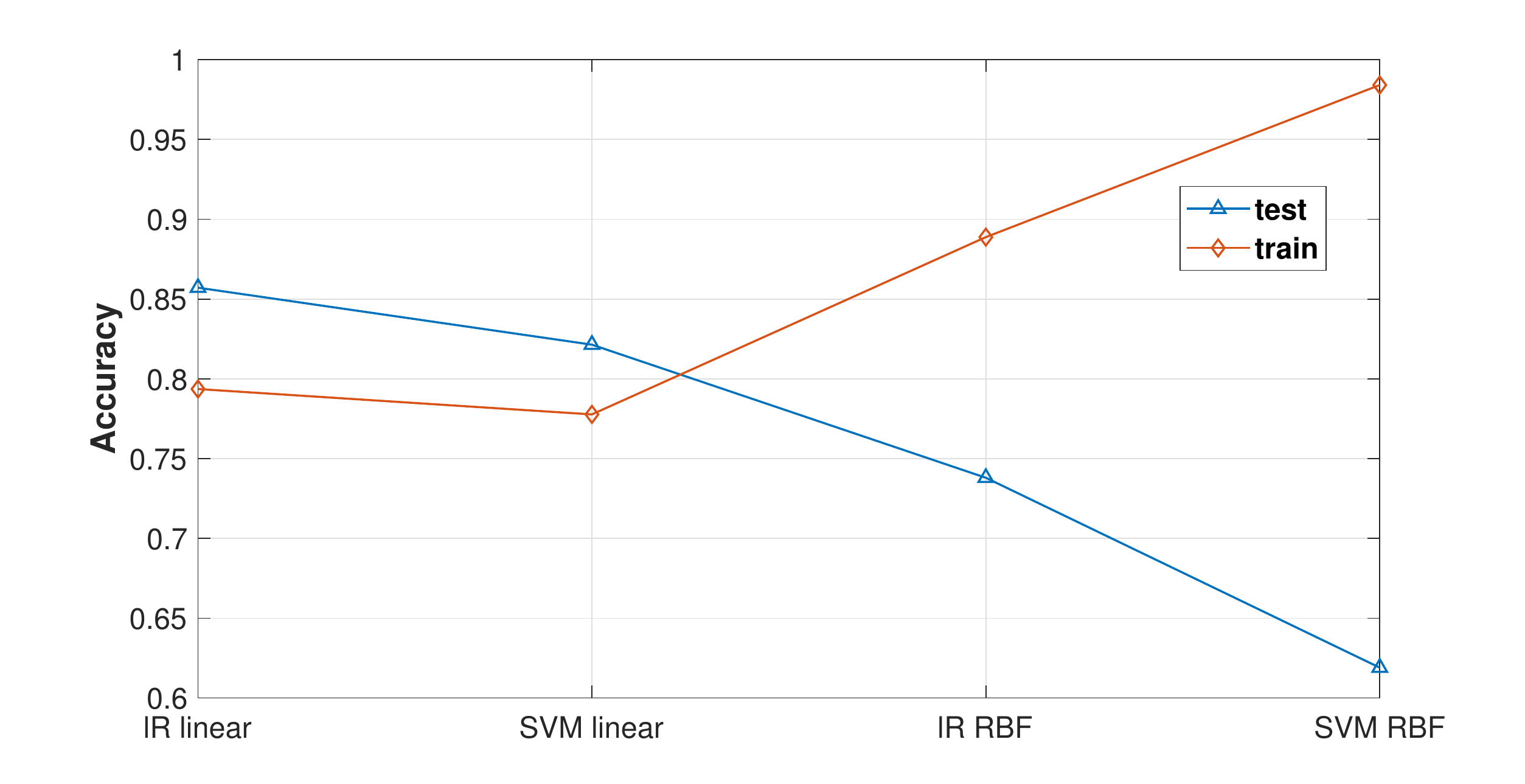}
\caption{}
\label{sfig:acc2}
\end{subfigure}
\caption{Testing and training accuracies for various classifiers arranged in the order of classification model complexity from left to right. The estimated accuracies for dataset-I and dataset-II are shown in (a) and (b) respectively.}
\label{fig:simAcc}
\end{figure}

\subsection{Classification Performance}

Finally, the performance of the classifiers using the features explained for both the datasets are shown in Figure\,\ref{fig:simAcc}. The classifiers are arranged in the order of complexity from left to right with logistic regression (lR) and linear kernel being simplest and SVM with RBF kernel being most complex. The performance plot parallels the classic machine learning divergence curve for both the datasets. The accuracy for training data increases when increasing the classification model complexity while it reduces for the testing data. This is intuitive because a complex classification model would try to better classify the training data. But the performance of the test data would reduce due to overfitting upon using the complex models. We have very few training examples to build the classifier and hence such trend is expected. The performance of the classifiers for both the datasets are fairly high which reflects the strength of the estimated features. We can see a $87.6\%$ test accuracy for dataset-I and $85.7\%$ for dataset-II. While these accuracies depend a lot on the cross-validation numbers and other factors like choice of classifier which can be better tuned to get higher numbers. 

For both the datasets we have seen that the proposed methodology efficiently extracts the features which serves as good candidate to differentiate the imagined motor movements. By implicitly removing the effects of the unwanted stimuli, the coefficients of the coupling matrix $A$ are shown to be sufficient for discriminating relation between various EEG signals which are indicative of the motor movements. The testing accuracies are high which indicate the good quality of the extracted features.

\section{Conclusion}
\label{sec:concl}

We have revisited the EEG-based noninvasive brain interfaces feature extraction and translation from a cyber-physical systems' lens. Specifically, we leveraged spatiotemporal fractional-order models that cope with the unknown inputs. The fractional-order models provide us the dynamic coupling changes that rule the EEG data collected from the different EEG sensors, and the fractional-order exponents capture the long-term memory of the process. Subsequently, unknown stimuli is determined as the external input that least conforms with the fractional-order model. By doing so, we have filtered-out from the brain EEG signals the unknown inputs, that might be originated in the deeper brain structures. The presence of unknown stimuli is possibly the result of the structural connectivity of the brain that crisscrosses different regions, or due to artifacts originated in the muscles (e.g., eye blinking or head movement). As a consequence, the filtered signal does not need to annihilate an entire band in the frequency domain, thus keeping information about some frequency regions of the signal that would be otherwise lost. 

We have shown how the different features obtained from the proposed model can be used towards rethinking the EEG-based noninvasive interfaces. In particular, two datasets used in BCI competitions were used to validate the performance of the methodology introduced in this manuscript, which is compatible with some of the state-of-the-art performances while requiring a relatively small number of training points. We believe that the proposed methodology can be used within the context of different neurophysiological processes and corresponding sensing technologies. Future research will focus on leveraging additional information from the unknown inputs retrieved to anticipate specific artifacts and enable the deployment of neuro-wearables in the context of real-life scenarios.  Furthermore, the presented methodology can be used as an exploratory tool by neuroscientists and physicians, by testing input and output responses and tracking their impact in the unknown inputs retrieved by the algorithm proposed; in other words, one will be able to systematically identify the origin and dynamics of stimulus across space and time. Finally, it would be interesting to explore the proposed approach in the closed-loop context, where the present models would benefit from control-like strategies to enhance the brain towards certain tasks or attenuate side effects of certain neurodegenerative diseases or disorders.

\section*{Acknowledgment}

The authors are thankful to the reviewers for their valuable feedback. G.G. and P.B. gratefully acknowledge the support by the U.S. Army Defense Advanced Research Projects Agency (DARPA) under grant no. W911NF-17-1-0076, DARPA Young Faculty Award under grant no. N66001-17-1-4044, and the National Science Foundation under CAREER Award CPS-1453860 support.

%
%

\footnotesize
\bibliographystyle{IEEEtran}
\bibliography{IEEEabrv,bciClassif}

\end{document}